\documentclass[12pt,a4paper]{article}

\usepackage{tikz}
\usetikzlibrary{arrows.meta,decorations.pathmorphing,decorations.markings}
\usepackage{amsmath,amssymb,amsthm}
\usepackage[normalem]{ulem}
\usepackage{physics}

\usepackage[margin=2.5cm]{geometry}
\usepackage[colorlinks=true,linkcolor=blue,citecolor=blue]{hyperref}
\usepackage{multirow}
\usepackage{tabularx}
\usepackage{booktabs}
\usepackage[table]{xcolor}
\usepackage{subcaption}
\usepackage{etoolbox}
\usepackage{comment}
\usepackage{multirow}
\usepackage{makecell}
\newcommand{\Pbar}{\overline{P}}
\newcommand{\Phat}{\widehat{P}}
\newcommand{\Pbarhat}{\widehat{\overline{P}}}

\begin{document}

\title{    \vspace*{-3cm}\makebox[\textwidth][r]{%
        \normalfont\normalsize{FERMILAB-PUB-26-0820-T}
    }
    \vspace{1cm}
    
Is Neutrino Disappearance Sensitive to T-Violation\\ for Asymmetric Matter Profiles ?}

\author{Gabriela Barenboim$^{a}$,
        Stephen J. Parke$^{b}$,
        Alberto S\'anchez Vargas$^{a}$}

\date{
  $^a$ Departament de F\'isica Te\`orica, Universitat de Val\`encia
       and IFIC, Universitat de Val\`encia--CSIC,
       E-46100 Burjassot, Spain\\[4pt]
  $^b$ Theoretical Physics Department, Fermi National Accelerator Laboratory,
       Batavia, IL 60510, USA\\[5mm]
       \today}
       
\maketitle

\begin{abstract}
It is commonly stated in the literature that T violation is zero in neutrino
disappearance channels.  We show this claim is not general: it holds
only for symmetric matter profiles or when the T/CP-violating phase ($\delta$) is 0 or $\pi$, that is, when the Jarlskog invariant is zero.
The $\nu_\mu$ and $\nu_\tau$, but not the $\nu_e$, survival probabilities carry a
$\sin\delta$-dependent term that is non-zero in asymmetric matter
and constitutes a genuine T asymmetry.  The confusion arises from
an improper T test that subtracts a probability from itself:
a mathematical identity, not a physics result. In this paper, we clarify the distinctions between a proper and an improper T test as well as the proper and improper CP and CPT tests, for a general, arbitrary matter profile. 

For a toy model consisting of two different layers of constant density, we find that the maximum T-violation is obtained when one layer is close to the solar resonance, and the other is close to the atmospheric resonance. We address whether the matter asymmetry for DUNE and Hyper-Kamiokande experiments is sufficiently large to allow  
the $\sin\delta$ dependence to be probed in $\nu_\mu$ survival probability. We also discuss future experimental scenarios that could probe T/CP-violation in $\nu_\mu$ survival probability.

\end{abstract}

\section{Introduction}
\label{sec:intro}

There are numerous papers in the literature that assumed that T violation is always zero
for neutrino disappearance channels.
However, there are a number of very early papers \cite{Kuo:1987km,Minakata:1999ze,Fishbane:2000kw} whose results imply that T-violation can be probed in disappearance for three flavor propagation in an asymmetric medium when the Jarlskog invariant is non-zero, e.g. $\delta\neq 0$ or $\pi$. The muon and tau, but not electron, neutrino survival probabilities can be used as genuine probes of T-violation in an asymmetric matter environment, as stated more explicitly later 
in Refs.~\cite{Akhmedov:2001kd, Yokomakura:2002av, Akhmedov:2004ve}.
The literature on T-violation in neutrino oscillation is vast, some other early works are Refs.~\cite{Cabibbo:1977nk,Krastev:1988yu,Toshev:1989vz,Toshev:1991ku, Arafune:1996bt}, but most of the literature focuses on symmetric matter profiles and only addresses T-violation in the appearance channels, comparing $\nu_\alpha \to \nu_\beta$ with $\nu_\beta \to \nu_\alpha$, for $\alpha \neq \beta$. We argue that for a proper T-violation comparison, a reversal of the direction of propagation is required for asymmetric matter profiles, as given in Refs.~\cite{Schwetz:2021thj, Kitano:2024kdv,Bitter:2024eka}.

However, none of the above referenced papers or any other papers address T-violation in asymmetric matter profiles in a comprehensive manner, especially for disappearance channels.  
In this paper, we revisit the claims of Refs.~\cite{Kuo:1987km,Minakata:1999ze,Fishbane:2000kw, Akhmedov:2001kd, Yokomakura:2002av, Akhmedov:2004ve}, address how the above confusion has arisen and provide a coherent and comprehensive overview of T-violation in an asymmetric medium while also providing new insight, especially for $\nu_\mu$ disappearance experiments. The most closely related papers to our work are Refs.~\cite{Yokomakura:2002av, Akhmedov:2004ve}. Nevertheless, these papers only address T-violation in the disappearance channels for asymmetric matter profiles briefly in passing. This is why T-violation in asymmetric matter for disappearance channels needs to be addressed correctly, coherently and comprehensively.
\\

This confusion stems from the commonly used  quantity to test T violation, which compares the probabilities for $\nu_\alpha \to \nu_\beta$ with $\nu_\beta \to \nu_\alpha$. We will call this definition as an \textbf{improper} T test. In disappearance channels, it compares the probability of $\nu_\alpha\to\nu_\alpha$ with itself, giving
zero by construction.
But this improper T test swaps initial and final flavors while leaving the
matter profile unchanged. This is only equivalent to a genuine
time reversal for symmetric profiles: for an asymmetric profile, a \textbf{proper}
time reversal must also reverse the matter traversed by the
neutrino, not just the flavor labels. That is equivalent to interchange neutrino source and detector: comparing $\nu_\alpha \to \nu_\beta$ with $\nu_\alpha \leftarrow \nu_\beta$. Once this is done
correctly, the proper T asymmetries in the $\nu_\mu$ and $\nu_\tau$ (but not $\nu_e$)
survival channels are non-zero for an asymmetric matter profile, given that the Jarlskog invariant~\cite{Jarlskog:1985ht} is non-vanishing. Some works apply correctly this proper time-reversal definition, but they either restricted to symmetric matter~\cite{Kitano:2024kdv} or did not arrive to the conclusion that disappearance channels are sensitive to T-violation~\cite{Schwetz:2021thj, Bitter:2024eka}. Other works considered asymmetric profiles but applied an improper test~\cite{Miura:2001pi, Miura:2001pia} or did not fully identify the forward/backward comparison as time-reversal~\cite{Akhmedov:2001kd, Akhmedov:2004ve}. \\

An analogous distinction applies to CP (for a detailed discussion see the reviews~\cite{Nunokawa:2007qh, Branco:2011zb} and references therein). The test commonly used
in experiments compares neutrinos and antineutrinos propagating
through the same matter, which contaminates the result with
matter-induced effects unrelated to genuine CP violation.  The \textbf{proper} CP test instead compares {\it neutrinos in matter} with
{\it antineutrinos in antimatter}.  
A central result of this paper is that the two proper T and CP tests
yield identical asymmetries for any matter
profile, and that both are purely sensitive to genuine T/CP
violation in the mixing matrix, with no contamination from the
propagation medium. The \underline{intrinsic} contribution
to an asymmetry is proportional to the Jarlskog invariant ($\sin\delta$); whereas an
\underline{extrinsic} contribution can be non-zero even when the Jarlskog invariant is zero ($\delta=0,\pi$). As we will show later, proper tests receive only intrinsic contributions.\\

The paper is organised as follows: Section~\ref{sec:PandImp} discusses the proper and improper T and CP tests as well as CPT and relationships between them.
CPT invariance imposes an identity relating the oscillation probabilities. We call this the reversal identity, which relates the probabilities forward in a matter profile to those backwards in the same matter profile.
In Sec.~\ref{sec:general probabilities} we re-derive the form of the oscillation probabilities in a general  asymmetric profile and its implication for the proper and improper T and CP tests. Sec.~\ref{sec:2 layers} illustrates the physics of an asymmetric matter profile through two layers with different constant densities. The answer to ``Are DUNE~\cite{DUNE:2020ypp} and Hyper-Kamiokande (HK)~\cite{Hyper-Kamiokande:2018ofw} sensitive to these effects?'' is given in Sec.~\ref{sec:DUNE+HyperK}. In Sec.~\ref{sec:futuristic experiments} we discuss two futuristic scenarios that demonstrate that T/CP-violation is measurable in $\nu_{\mu}$ disappearance. The conclusions are then given followed by a number of Appendices where technical details are addressed. 

%====================================================================
\section{Proper and Improper T and CP Transformations}
\label{sec:PandImp}
%====================================================================

Let us first define the three neutrino oscillation probability in matter, for $\nu_\alpha(x_0) \rightarrow \nu_\beta(x_1)$, to be
\begin{equation}
\label{eq:prob}
    P\,^\rightarrow_{\alpha\beta}[U,W],
\end{equation}
where $U$ is the Pontecorvo–Maki–Nakagawa–Sakata
(PMNS) matrix and $W(x)=\sqrt{2}G_FN_{e}(x)$
is the Wolfenstein matter potential \cite{Wolfenstein:1977ue} with $N_e$ the electron (positron) number density for matter (anti-matter).
The  ``$\rightarrow$'' superscript indicates that the propagation is from $x_0 \to x_1$ in the matter potential. Backwards propagation is the process $\nu_\alpha(x_0) \leftarrow  \nu_\beta(x_1) $, for which we define the following probability 
\begin{equation}
    P\,^\leftarrow_{\alpha \beta }[U,W].
\end{equation}
The transformation that goes from forward to backward propagation is time reversal, T.\\

We define analogous oscillation probabilities for anti-neutrinos, also when the propagation medium is composed of antimatter. All these quantities can be related to eq.~\eqref{eq:prob} with appropriate arguments:
\begin{subequations}
\label{eq:Ps}
\begin{align}
\label{eq:Pbar}
 \text{$\overline{\nu}$ in matter:} \quad  &\Pbar\,^\rightarrow_{\alpha\beta}[U,W] = P\,^\rightarrow_{\alpha\beta}[U^*,-W] 
 \\
\label{eq:Phat}
 \text{$\nu$ in antimatter:} \quad   &\Phat\,^\rightarrow_{\alpha\beta}[U,W] = P\,^\rightarrow_{\alpha\beta}[U\phantom{^*},-W] 
    \\
    \label{eq:Pbarhat}
  \text{$\overline{\nu}$ in antimatter:} \quad &\Pbarhat\,^\rightarrow_{\alpha\beta}[U,W] = P\,^\rightarrow_{\alpha\beta}[U^*,\phantom{-}W] 
\end{align}    
\end{subequations}
and similarly for the backward propagation probabilities. Changing neutrinos to anti-neutrinos, one replaces $U \rightarrow U^*$ and flips the sign of W, whereas going from matter to anti-matter one just flips the sign of W. When considering time reversal, it is sometimes useful to introduce the reversed matter profile given by
$W_{\rm rev}(x)\equiv W(x_0+x_1-x)$,
\begin{align}
\label{eq:PWrev}
\hspace*{-1.45cm}\text{$\nu$ backwards in matter:} \quad &P\,^\leftarrow_{\alpha\beta}[U,W]
 = P\,^\rightarrow_{\beta\alpha}[U,W_{\rm rev}]
\end{align} 
and a matter profile is symmetric if $W_{\rm rev} = W$. \\

There is a set of transformations that relate all these probabilities between them. Departing from the oscillation probability for neutrinos forward in matter, $P\,^\rightarrow_{\alpha\beta}[U,W]$,
the action of the operators $\{ 1, \mathrm{cp}, \tau, \mathrm{cp}\tau, \mathrm{CP}, \mathrm{T}, \mathrm{CPT}\}$, and their physical interpretation can be stated as:

\begin{equation}
\begin{aligned}
&\textbf{Unity} \\
\mathrm{1}:   \quad & P\,^\rightarrow_{\alpha\beta}[U,W]
                      \to P\,^\rightarrow_{\alpha\beta}[U,W]  \quad ~~ [\nu_\alpha \to \nu_\beta \text{ forward in matter}]\\[6 pt]
&\textbf{Improper transformations} \\
\mathrm{cp}: \quad & P\,^\rightarrow_{\alpha\beta}[U,W]
                      \to \overline{P}\,^\rightarrow_{\alpha\beta}[U,W] \quad ~~ [\overline{\nu}_\alpha \to \overline{\nu}_\beta\text{ forward in matter}]\\
\tau: \quad & P\,^\rightarrow_{\alpha\beta}[U,W]
                      \to P\,^\rightarrow_{\beta\alpha}[U,W]  \quad ~~ [\nu_\beta \to \nu_\alpha\text{ forward in matter}]\\
\mathrm{cp}\tau: \quad & P\,^\rightarrow_{\alpha\beta}[U,W]
                      \to \overline{P}\,^\rightarrow_{\beta\alpha}[U,W]  \quad ~~ [\overline{\nu}_\beta \to \overline{\nu}_\alpha\text{ forward in matter}] \\[6 pt]
&\textbf{Proper transformations} \\                      
\mathrm{CP}:  \quad & P\,^\rightarrow_{\alpha\beta}[U,W]
                      \to \widehat{\overline{P}}\,^\rightarrow_{\alpha\beta}[U,W] \quad ~~ [\overline{\nu}_\alpha \to \overline{\nu}_\beta\text{ forward in antimatter}]\\
\mathrm{T}:  \quad & P\,^\rightarrow_{\alpha\beta}[U,W]
                      \to P\,^\leftarrow_{\alpha\beta}[U,W] \quad ~~ [\nu_\beta \to \nu_\alpha\text{ backwards in matter}]\\
\mathrm{CPT}: \quad & P\,^\rightarrow_{\alpha\beta}[U,W]
                      \to \widehat{\overline{P}}\,^\leftarrow_{\alpha\beta}[U,W]  \quad ~~ [\overline{\nu}_\beta \to \overline{\nu}_\alpha\text{ backwards in antimatter}]\\
\label{eq:operators}
\end{aligned}
\end{equation}
We denote $\{\mathrm{CP}, \mathrm{T}, \mathrm{CPT}\}$ as \textbf{proper} transformations since they transform both the neutrino state and its propagation environment; whereas $\{\mathrm{cp}, \tau, \mathrm{cp}\tau\}$ are \textbf{improper} ones, as they only transform the neutrino state. For example, a proper CP transformation maps matter into antimatter and neutrinos into antineutrinos, while a proper time reversal inverts the propagation direction through the same matter profile, as well as swapping initial and final neutrino flavor. Transforming only the neutrino state while leaving the external environment unchanged defines a different operation; that is what we refer to as an improper transformation. However, these are the commonly used transformations in the literature, since they can be easily probed in experiments.\\

If CPT is conserved, then $\widehat{\overline{P}}\,^\leftarrow_{\alpha\beta}[U,W] = P\,^\rightarrow_{\alpha\beta}[U,W]$, or equivalently, using the backwards counterpart of eq.~\eqref{eq:Pbarhat}, we arrive to  the {\bf Reversal Identity} for probabilities:
\begin{equation}
\label{eq:master}
    P\,^\leftarrow_{\alpha\beta}[U^*,W] = P\,^\rightarrow_{\alpha\beta}[U,W],
\end{equation}
which equals $P\,^\rightarrow_{\beta\alpha}[U^*,W_{\rm rev}]$, using the definition $W_{\rm rev}$ presented in eq.~\eqref{eq:PWrev}. The reversal identity, first derived and discussed in Refs.~\cite{Akhmedov:2001kd} and \cite{Akhmedov:2004ve} means that backwards propagation can always be mapped to forward propagation using $U \to U^*$. This single identity is important for all proper and improper transformations.
The three conditions for it to hold are: linearity of the evolution
equation, hermiticity of $H$ and a real\footnote{Note that a complex matter potential, $W \neq W^*$, a possibility in BSM models like neutrino nonstandard interactions (NSI), would just generalize eq.~\eqref{eq:master} into $P\,^\leftarrow_{\alpha\beta}[U,W]=
P\,^\rightarrow_{\alpha\beta}[U^*,W^*]$. } $W(x)$. Further discussion of the reversal identity is given in Appendix~\ref{app:reverse}. \\ 

We highlight that the process $\nu_\beta \to \nu_\alpha$ forward in matter is not the same as $\nu_\beta \to \nu_\alpha$ backwards in matter, at least for asymmetric matter profiles. In fact, the T test confusion stems from conflating the proper time reversal, $\mathrm{T}$, that inverts the propagation direction, with the improper $\tau$, that just swaps initial and final flavor. This confusion is reasonable, since for symmetric matter profiles both transformations have the same effects on probabilities.
Then, the commonly used T probability difference in the literature is actually what we call the improper $\tau$ test
\begin{equation}
\label{eq:fake T}
\Delta P^{\,\tau}_{\alpha\beta}
\equiv
P\,^\rightarrow_{\alpha\beta}[U,W]
-
P\,^\rightarrow_{\beta \alpha}[U,W]\,.
\end{equation}
\\
Setting $\alpha=\beta$ immediately
gives $\Delta P^{\,\tau}_{\alpha\alpha}=0$: the expression subtracts
a probability from itself. But this is only a correct or proper T
test for symmetric matter profiles.  For an asymmetric profile,
the correct or proper T test compares forward propagation with backward
propagation through the same matter profile~\cite{Schwetz:2021thj,Bitter:2024eka}:

\begin{equation}
\begin{aligned}
\label{eq:true T}
\Delta P^{\rm T}_{\alpha\beta}
&\equiv
P\,^\rightarrow_{\alpha\beta}[U,W]
-
P\,^\leftarrow_{\alpha\beta}[U,W] \\
&=P\,^\rightarrow_{\alpha\beta}[U,W]
-
P\,^\rightarrow_{\alpha\beta}[U^*,W],
\end{aligned}
\end{equation}
where we have used eq.~\eqref{eq:master} in the second equality. That is, $\Delta P^{\rm T}_{\alpha\beta}$ measures the part of the probability that is odd under $U\to U^*$. As we will show, this is non-zero even for
$\alpha= \beta =\mu,\tau$ in an asymmetric matter profile, when the Jarlskog invariant is non-zero\footnote{This requires the Dirac CP phase $\delta\neq 0,\pi$ and all the mixing angles, $\theta_{ij}\neq 0,\pi$ for the PDG parameterization of $U$~\cite{ParticleDataGroup:2020ssz}.}. The electron disappearance channel is special:
$\Delta P^{\rm T}_{ee}=0$ for any matter profile because  the $(ee)$ component of the Hamiltonian has a special role for the Wolfenstein matter potential, as we will explain in more detail in Section~\ref{sec:general probabilities}. 

An analogous distinction applies to CP.  The improper cp test
compares {\it neutrinos in matter} with {\it anti-neutrinos in matter}.
Whereas the the proper CP test compares {\it neutrinos in matter} with {\it anti-neutrinos in anti-matter}. 
Then the improper cp test is given by
\begin{align}
\label{eq:false CP}
\Delta P^{\rm cp}_{\alpha\beta}
&\equiv
P\,^\rightarrow_{\alpha\beta}[U,W]
-
\overline{P}\,^\rightarrow_{\alpha\beta}[U,W]
=
P\,^\rightarrow_{\alpha\beta}[U,W]
-
P\,^\rightarrow_{\alpha\beta}[U^*,-W],
\end{align}
which in matter is non-zero even if the Jarlskog invariant is zero, that is, introduces extrinsic CP violation.\\

Whereas the proper CP test is given by
\begin{equation}
    \begin{aligned}
\label{eq:true CP}
\Delta P^{\rm CP}_{\alpha\beta}
&\equiv
P\,^\rightarrow_{\alpha\beta}[U,W]
-
\widehat{\overline{P}}\,^\rightarrow_{\alpha\beta}[U,W] \\
 &=
P\,^\rightarrow_{\alpha\beta}[U,W]
-
P\,^\rightarrow_{\alpha\beta}[U^*,W]\,,
\end{aligned}
\end{equation}
and vanishes iff the Jarlskog invariant is zero, e.g. when $\delta=0,\pi$. 
Thus,
\begin{equation}
\label{eq:T equal CP}
\Delta P^{\rm T}_{\alpha\beta}
=\Delta P^{\rm CP}_{\alpha\beta},
\end{equation}
for any matter profile.
The equivalence between the proper T and proper CP tests follows directly from the exact relation between
backward propagation and complex conjugation of the mixing matrix, given by the reversal identity in eq.~\eqref{eq:master}, or in other words, from CPT invariance. Explicitly,
\begin{equation}
    \begin{aligned}
\label{eq:true CPT}
\Delta P^{\rm CPT}_{\alpha\beta}
&\equiv
P\,^\rightarrow_{\alpha\beta}[U,W]
-
\widehat{\overline{P}}\,^\leftarrow_{\alpha\beta}[U,W] \\
 &= P\,^\rightarrow_{\alpha\beta}[U,W]
-
P\,^\leftarrow_{\alpha\beta}[U^*,W] \\
&= 0.
\end{aligned}
\end{equation}
This equation had already been pointed out, in a different formulation, in Ref.~\cite{Akhmedov:2004ve}. \\

In Table \ref{tab:symmetries}, we summarize the relationships between the different operations on the three possible matter profiles: vacuum, symmetric and asymmetric.

\begin{table}[htb!]
\centering
\begin{tabular}{c |c c c}
\hline
Matter Profile: & Vacuum & Symmetric & Asymmetric \\
\hline

\multirow{3}{*}{\textbf{Appearance:}}
& $\tau = \mathrm{T}$
& $\tau = \mathrm{T}$
& $\tau \neq \mathrm{T}$ \\[4pt]

& $\mathrm{cp} = \mathrm{CP}$
& $\mathrm{cp} \neq \mathrm{CP}$
& $\mathrm{cp} \neq \mathrm{CP}$ \\[4pt]

& $\mathrm{cp}\tau = \mathrm{CPT}$
& $\mathrm{cp}\tau \neq \mathrm{CPT}$
& $\mathrm{cp}\tau \neq \mathrm{CPT}$ \\[2mm]

\hline
\hline\\[-2mm]
\multirow{3}{*}{\textbf{Disappearance:}}
&  $\tau = 1 = \mathrm{T}$
& $\tau = 1 = \mathrm{T} $
& $\tau = 1 \neq \mathrm{T}$ \\[4pt]

& $\mathrm{cp} = \mathrm{CP}$
& $\mathrm{cp} \neq \mathrm{CP}$
& $\mathrm{cp} \neq \mathrm{CP}$ \\[4pt]

& $%\mathrm{cp} = 
\mathrm{cp}\tau = \mathrm{CPT}$
& $%\mathrm{cp} =
\mathrm{cp}\tau \neq \mathrm{CPT}$
& $%\mathrm{cp} = 
\mathrm{cp}\tau \neq \mathrm{CPT}$ \\[2mm]

\hline
\end{tabular}
\caption{Relation between improper and proper transformations in different propagation media: vacuum, symmetric matter or asymmetric matter. For disappearance, $\tau =1$, since swapping initial and final flavor has no effect. However, for asymmetric matter profiles, $T \neq 1$ for $\nu_\mu$ and $\nu_\tau$ disappearance  but $T= 1$ for $\nu_e$  disappearance. $\mathrm{cp}\tau$ and CPT can only have the same effect in vacuum.}
\label{tab:symmetries}
\end{table}

\paragraph{Probability Diagrams:}

\begin{figure}
    \centering
    \includegraphics[width=\linewidth]{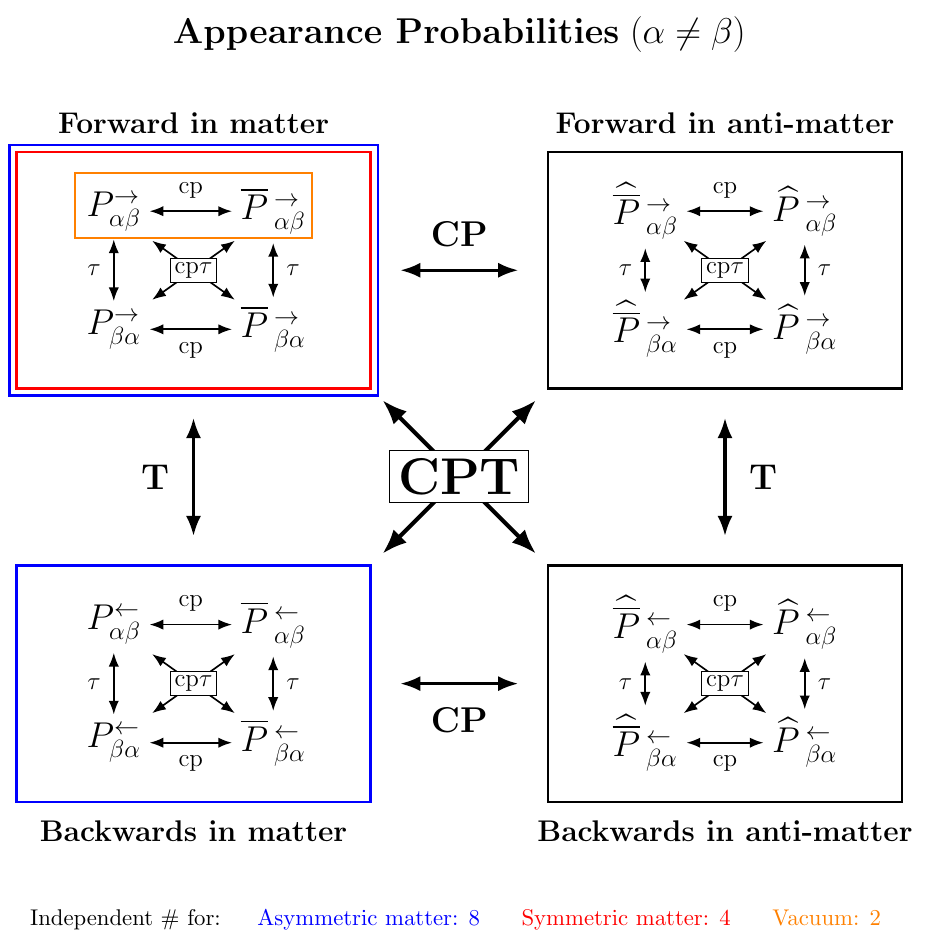}
    \caption{For a generic matter profile, starting from the appearance probability $P\,^\rightarrow_{\alpha\beta}[U,W]$ in the left top corner, the other 3 probabilities in the bigger blue box are related by cp, $\tau$ and cp$\tau$. Using CP, T and CPT on this cluster, one can relate to the other 3 probability clusters. The two blue boxes give the  probabilities in matter and the same probabilities in anti-matter can be obtained by CPT. The bottom and top blue boxes of probabilities are related by $U \to U^*$, as this implements the T operation: the interchange of forward to backwards propagation. If the matter profile is symmetric, then $\tau$=T, so that $P^\leftarrow_{\beta \alpha}=P^\rightarrow_{\alpha\beta}$, etc, thus one only needs the probabilities in the red box with others obtained by CPT. Again the second row of this box can be obtained from the first by replacing U with U$^*$.  In vacuum, $\tau$=T, cp=CP and therefore cp$\tau$ = CPT and only the two probabilities in the orange, smallest box are needed.}
    \label{fig:app}
\end{figure}

In  Fig. \ref{fig:app}, we present all relations between the different appearance oscillation probabilities,  that correspond to the physical processes of neutrinos/antineutrinos of initial and final flavor ($\alpha$, $\beta$) or ($\beta$, $\alpha$), with $\alpha \neq \beta$, propagating in a medium composed of matter (left boxes) or antimatter (right boxes), propagating forward (top boxes) or backwards (bottom boxes) in the matter profile. These 4 situations yield 16 different oscillation probabilities. Since CPT is an exact symmetry, the probabilities in the left top box are equal to the probabilities in the right bottom box and similarly for right top with left bottom, so that there are only $8$ independent probabilities, the probabilities in the blue (or black) boxes. The probabilities in the bottom boxes are related to the ones in the top boxes by T, and the reversal identity given in eq.~\eqref{eq:master} connects backwards propagation to forward propagation by replacing $U \to U^*$. Similarly, the probabilities in the right boxes are related to the ones in the left boxes by CP, which also amounts to replacing $U \to U^*$. The boxes across the diagonal are then related by CPT, which is equivalent to apply
$U \to U^*$ twice, and leaves the probabilities unchanged.
\\

When restricting ourselves to symmetric matter profiles, the operations $\tau$ and T are identical and the probabilities in the bottom boxes are contained in the top boxes, therefore, we only need the 4 probabilities in the left top (red) box. If we further consider vacuum, we just need the top two probabilities in the left top (orange) box as cp$\tau$ is equivalent to CPT.\\

\begin{figure}
    \centering
    \includegraphics[width=\linewidth]{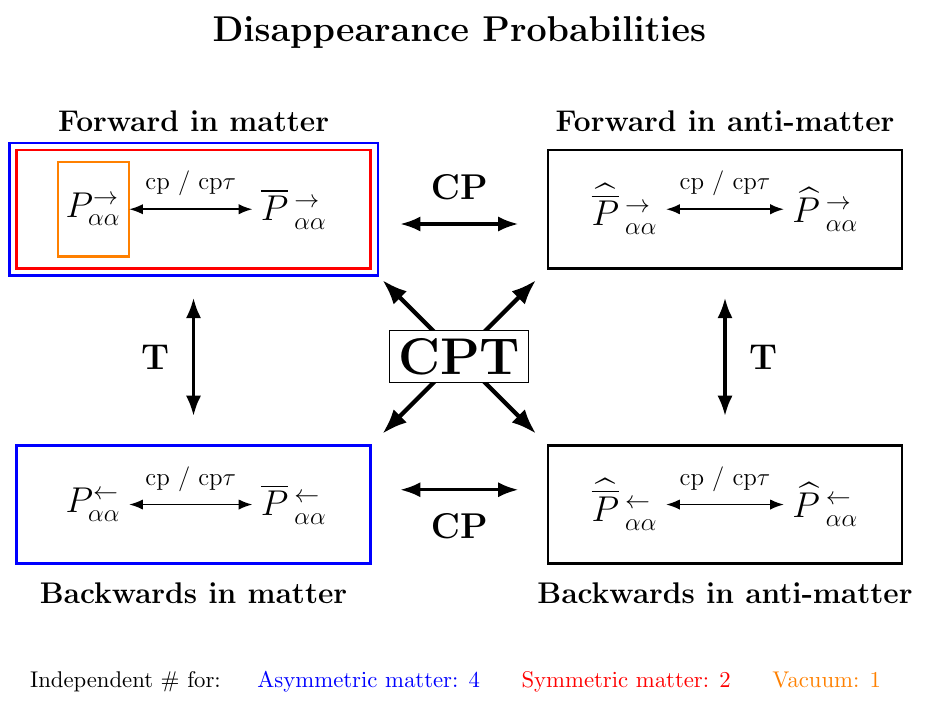}
    \caption{For a generic matter profile, starting from the disappearance probability $P\,^\rightarrow_{\alpha\alpha}$ in the top left hand corner, the other probability in the smaller probability clusters is obtained by cp or cp$\tau$ (since $\tau = 1$ for disappearance). Using CP, T and CPT on this cluster, one can relate to the other 3 probability clusters. The two blue boxes give the  probabilities in matter and the same probabilities in anti-matter can be obtained by CPT. In general, we will show that $P^\leftarrow_{\alpha \alpha} \neq P^\rightarrow_{\alpha\alpha}$, for $\alpha =\mu \rm{ ~or~ } \tau$, for an asymmetric matter profile. These two probabilities are related by $U \to U^*$. Electron neutrino disappearance is special in three flavors, as $P^\leftarrow_{ee} = P^\rightarrow_{ee}$ for all matter profiles because of the form of the Wolfenstein matter potential. If the matter profile is symmetric, for the disappearance channels, $\tau$=T=1, so that $P^\leftarrow_{\alpha \alpha}=P^\rightarrow_{\alpha\alpha}$, thus one only needs the probabilities in the red box with others obtained by CPT. In vacuum, for disappearance, $\tau$=T=1, cp=CP and therefore cp=cp$\tau$=CPT and only the one probability in the orange, smallest box is needed. 
}
    \label{fig:dis}
\end{figure}

In  Fig.~\ref{fig:dis}, we present all relations between the different disappearance oscillation probabilities, that correspond to the physical processes of neutrinos/antineutrinos of initial and final flavor ($\alpha$), propagating in a medium composed of matter (left boxes) or antimatter (right boxes), propagating forward (top boxes) or backwards (bottom boxes) in the matter profile. A similar discussion to the one for appearance channels holds for the disappearance channels except that the number of probabilities in each box is halved as the improper T operation ($\tau$) is equal to the identity operation. \\

In summary, proper tests transforms both the neutrino state and its
environment and improper tests transform only the neutrino state, and are related to experimentally accessible quantities.
Table~\ref{tab:proper_improper} collects all six transformations.

\begin{table}[h]
\centering
\renewcommand{\arraystretch}{1.8}
\begin{tabular}{c@{\hspace{1cm}}r@{\hspace{0.15cm}}c@{\hspace{0.15cm}}l}
\hline\hline
Tests & \multicolumn{3}{c}{Definition} \\
\hline

\multirow{2}{*}{\makecell{T\\ (proper)}}
  & $\Delta P^{\rm T}_{\alpha\beta}$
  & $\equiv$
  & $P\,^\rightarrow_{\alpha\beta}[U,W]
     - P\,^\leftarrow_{\alpha\beta}[U,W]$ \\
  &
  & $=$
  & $P\,^\rightarrow_{\alpha\beta}[U,W]
     - P\,^\rightarrow_{\alpha\beta}[U^*,W]$ \\

\makecell{$\tau$\\ (improper)}
  & $\Delta P^{\tau}_{\alpha\beta}$
  & $\equiv$
  & $P\,^\rightarrow_{\alpha\beta}[U,W]
     - P\,^\rightarrow_{\beta\alpha}[U,W]$ \\[10pt]

\hline

\multirow{2}{*}{\makecell{CP\\ (proper)}}
  & $\Delta P^{\rm CP}_{\alpha\beta}$
  & $\equiv$
  & $P\,^\rightarrow_{\alpha\beta}[U,W]
     - \Pbarhat\,^\rightarrow_{\alpha\beta}[U,W]$ \\
  &
  & $=$
  & $P\,^\rightarrow_{\alpha\beta}[U,W]
     - P\,^\rightarrow_{\alpha\beta}[U^*,W]$ \\

\multirow{2}{*}{\makecell{cp\\ (improper)}}
  & $\Delta P^{\rm cp}_{\alpha\beta}$
  & $\equiv$
  & $P\,^\rightarrow_{\alpha\beta}[U,W]
     - \overline{P}\,^\rightarrow_{\alpha\beta}[U,W]$ \\
  &
  & $=$
  & $P\,^\rightarrow_{\alpha\beta}[U,W]
     - P\,^\rightarrow_{\alpha\beta}[U^*,-W]$ \\[2pt]

\hline

\multirow{2}{*}{\makecell{CPT\\ (proper)}}
  & $\Delta P^{\rm CPT}_{\alpha\beta}$
  & $\equiv$
  & $P\,^\rightarrow_{\alpha\beta}[U,W]
     - \Pbarhat\,^\leftarrow_{\alpha\beta}[U,W]$ \\
  &
  & $=$
  & $P\,^\rightarrow_{\alpha\beta}[U,W]
     - P\,^\leftarrow_{\alpha\beta}[U^*,W]=0$ \\

\multirow{2}{*}{\makecell{cp$\tau$\\ (improper)}}
  & $\Delta P^{\rm cp\tau}_{\alpha\beta}$
  & $\equiv$
  & $P\,^\rightarrow_{\alpha\beta}[U,W]
     - \overline{P}\,^\rightarrow_{\beta\alpha}[U,W]$ \\
  &
  & $=$
  & $P\,^\rightarrow_{\alpha\beta}[U,W]
     - P\,^\rightarrow_{\beta\alpha}[U^*,-W]\neq 0$ \\

\hline\hline
\end{tabular}

\caption{Probability differences for proper (T, CP, CPT) and improper tests ($\tau$, cp, cp$\tau$). The
equivalent forms use eqs.~\eqref{eq:Ps} and CPT conservation,
encapsulated in the reversal identity, eq.~\eqref{eq:master}. As a consequence, both proper T
and proper CP reduce to $P[U,W]-P[U^*,W]$, so
$\Delta P^{\rm T}=\Delta P^{\rm CP}$ for any channel. The proper CPT
test is identically zero, whereas the improper cp$\tau$ is only zero
in vacuum. Furthermore, the improper T test ($\tau$) is identically
zero for disappearance channels independent of the form of the matter
profile. However, the proper T test can be non-zero even for the
disappearance channels.}
\label{tab:proper_improper}
\end{table}
\newpage

%====================================================================
\section{General form of three-flavor neutrino probabilities}
\label{sec:general probabilities}
%====================================================================
Three-flavor neutrino oscillation probabilities are given by
\begin{equation}
\label{eq:osc prob}
P\,^\rightarrow_{\alpha \beta}\bigl[U,W\bigr]
= \bigl|S_{\beta\alpha}(x,x_0,U,W)\bigr|^2,
% = \bigl|S^*_{\beta\alpha}(x_1,x_0,U,W)\bigr|^2.
\end{equation}
where $S(x,x_0, U, W)$ governs the evolution of neutrino flavors, $\nu=(\nu_e,\nu_\mu,\nu_\tau)^T$, from an initial state $\nu(x_0)$, $$\nu(x)=S(x,x_0,U,W)\,\nu(x_0).$$
The evolution matrix is obtained solving the Schr\"odinger-like equation,
\begin{equation}
i\frac{d}{dx}S(x,x_0,U,W)= H(x)\,S(x,x_0,U,W),
\end{equation}
where the Hamiltonian in matter, written in the flavor basis, reads as 
\begin{equation}
\label{eq:Hmat}
H(x) = \frac{1}{2E_\nu}\,UM^2U^\dagger
+ W(x)\begin{pmatrix}1&0&0\\0&0&0\\0&0&0\end{pmatrix},
\quad 
\end{equation}
where  $M^2=\mathrm{diag}(0,\Delta m^2_{21},\Delta m^2_{31})$ and the Wolfenstein matter potential is given by 
$W(x)=\sqrt{2}G_FN_{e}(x)$. $U$ is the PMNS matrix, which we will parametrize as 
\begin{align}
U_\text{PMNS} &= R_{23}(\theta_{23})D_3(\delta) R_{13}(\theta_{13}) R_{12}(\theta_{12})=U_\text{PDG} D_3(\delta)
\label{eq:UPDG}
\end{align}
where $\delta$ is the Dirac CP phase, $R_{ij}$ the orthogonal rotation matrices with angle $\theta_{ij}$, and $D_3(\delta) = \mathrm{diag}(1, 1, e^{i\delta})$. $U_{\rm PDG}$ is written in the same notation as in Ref.~\cite{ParticleDataGroup:2020ssz}. We use the shorthand notation $c_{ij} \equiv \cos\theta_{ij}$, $s_{ij} \equiv \sin\theta_{ij}$.
\\

This parametrization of the PMNS matrix is most useful for propagation through matter since
$ R_{23}(\theta_{23})D_3(\delta)$ commutes with the Wolfenstein matter potential, $W$, and the remainder of the PMNS matrix, $R_{13}(\theta_{13}) R_{12}(\theta_{12})$ is real.  Therefore, one can write the explicit form of the oscillation probabilities separating matter effects from the dependence on the 23 rotation and on the CP phase~\cite{Yokomakura:2002av, Blennow:2013rca, Ge:2013zua, Denton:2025oxf},
and in particular one can separate the odd and even parts under $U \to U^*$. Thus, we work in a rotated flavor basis, where
\begin{equation}
\label{eq:rotation}
   \widetilde{\nu}=(R_{23}D_3)^\dagger\nu.
\end{equation}
and the oscillation amplitudes are given by $\widetilde{S}$. Then, the full evolution matrix in the flavor basis, $S$, reads as
\begin{align}
    S &= R_{23}D_3 ~\widetilde{S} ~ D^*_3 R^T_{23}=
    \label{eq:Smatrix}
    \end{align}
    {\tiny
\begin{align}
    % &= \left(\begin{array}{ccc}
    % \widetilde{S}_{e e}  
    % & \widetilde{S}_{e \mu } c_{23}+\widetilde{S}_{e \tau } e^{-i\delta} s_{23}  
    % & \widetilde{S}_{ e \tau }  e^{-i\delta} c_{23}-\widetilde{S}_{e \mu} s_{23}\\[3mm] 
    % %
    % \widetilde{S}_{\mu e} c_{23}+\widetilde{S}_{\tau e}  e^{i\delta}  s_{23} 
    % &\widetilde{S}_{\mu \mu} c^2_{23} + \widetilde{S}_{\tau \tau}s^2_{23}+(\widetilde{S}_{\mu \tau}e^{-i\delta} + \widetilde{S}_{\tau \mu}e^{i\delta})c_{23}s_{23}
    % & \widetilde{S}_{\mu \tau} e^{-i\delta}c^2_{23}  - \widetilde{S}_{\tau \mu}  e^{i\delta} s^2_{23} 
    % +(\widetilde{S}_{\tau \tau}-\widetilde{S}_{\mu \mu} )c_{23}s_{23}\\[3mm]
    % %
    % \widetilde{S}_{\tau e}  e^{i\delta} c_{23}-\widetilde{S}_{\mu e} s_{23}
    % &  \widetilde{S}_{\tau \mu} e^{i\delta} c^2_{23} - \widetilde{S}_{\mu \tau}  e^{-i\delta} s^2_{23}  +(\widetilde{S}_{\tau \tau}-\widetilde{S}_{\mu \mu} )c_{23}s_{23} 
    % &\widetilde{S}_{\tau \tau}c^2_{23}+\widetilde{S}_{\mu \mu} s^2_{23} - (\widetilde{S}_{\mu \tau} e^{-i\delta} + \widetilde{S}_{\tau \mu}  e^{i\delta})c_{23}s_{23}
    % \end{array} \right) \,.
    % \\
    &\left(\begin{array}{ccc}
    \widetilde{S}_{11}  
    & \widetilde{S}_{12 } c_{23}+\widetilde{S}_{13 } e^{-i\delta} s_{23}  
    & \widetilde{S}_{ 13 }  e^{-i\delta} c_{23}-\widetilde{S}_{12} s_{23}\\[3mm] 
    \widetilde{S}_{21} c_{23}+\widetilde{S}_{31}  e^{i\delta}  s_{23} 
    &\widetilde{S}_{22} c^2_{23} + \widetilde{S}_{33}s^2_{23}+(\widetilde{S}_{23}e^{-i\delta} + \widetilde{S}_{32}e^{i\delta})c_{23}s_{23}
    & \widetilde{S}_{23} e^{-i\delta}c^2_{23}  - \widetilde{S}_{32}  e^{i\delta} s^2_{23} 
    +(\widetilde{S}_{33}-\widetilde{S}_{22} )c_{23}s_{23}\\[3mm]
    \widetilde{S}_{31}  e^{i\delta} c_{23}-\widetilde{S}_{21} s_{23}
    &  \widetilde{S}_{32} e^{i\delta} c^2_{23} - \widetilde{S}_{23}  e^{-i\delta} s^2_{23}  +(\widetilde{S}_{33}-\widetilde{S}_{22} )c_{23}s_{23} 
    &\widetilde{S}_{33}c^2_{23}+\widetilde{S}_{22} s^2_{23} - (\widetilde{S}_{23} e^{-i\delta} + \widetilde{S}_{32}  e^{i\delta})c_{23}s_{23}
    \end{array} \right) \, , \notag \end{align}}where the elements of $\widetilde{S}$ are only functions of the parameters $(\theta_{12},\Delta m^2_{21})$, $(\theta_{13}, \Delta m^2_{31})$ as well as the propagation length, the neutrino energy and the matter density, in the combinations $\{L/E_\nu,\rho E_\nu\}$. $\widetilde{S}$ is a unitary matrix with complex entries, $\widetilde{S}\widetilde{S}^\dagger =\widetilde{S}^\dagger \widetilde{S}=1$, and it is symmetric only for symmetric matter profiles. The terms below the diagonal can be obtained from those above the diagonal by interchanging the subscripts on $\widetilde{S}$ and changing the sign of $\delta$. On $\widetilde{S}$, we use the subscripts $(1,2,3)$ for the rotated flavor basis, $\widetilde{\nu}$, instead of the notation $(e,\mu^\prime, \tau^\prime)$, which is sometimes used. \\

Then, the general form of the probabilities is 
\begin{equation}
\begin{aligned}
\label{eq:P_alpha beta}
P\,^\rightarrow_{\alpha\beta}[U,W]
&= A_{\alpha\beta}
+ B_{\alpha\beta}\sin\delta
+ C_{\alpha\beta}\cos\delta
+ D_{\alpha\beta}\sin 2\delta
+ E_{\alpha\beta}\cos 2\delta,
\end{aligned}
\end{equation}
with real coefficients\footnote{Note that our notation differs from the one in Ref.~\cite{Yokomakura:2002av} for the labeling of the coefficients. Independently of the parameterization of $U$, $A_{\alpha\beta}+ C_{\alpha\beta}\cos\delta+ E_{\alpha\beta}\cos 2\delta $ is the even part and   $B_{\alpha\beta}\sin\delta + D_{\alpha\beta}\sin 2\delta $ is the odd part of $P\,^\rightarrow_{\alpha\beta}[U,W]$ under $U \to U^*$.  } 
$X_{\alpha\beta} \in (A,B,C,D,E)_{\alpha\beta}$ independent of $\delta$.
Only the channels not involving $\nu_e$ have all of these coefficients non-zero (explicit expressions can be found in Appendix~\ref{app:coefficients}).   Using eq.~\eqref{eq:Ps} we can easily construct the probability for antineutrinos in antimatter as,
$\Pbarhat\,^\rightarrow_{\alpha\beta}[U,W]
=P\,^\rightarrow_{\alpha\beta}[U^*,W] $, which changes the sign of $B$ and $D$ in eq.~\eqref{eq:P_alpha beta}. \\

The proper tests (both for proper T and proper CP) can be written explicitly as
\begin{equation}
\begin{aligned}
\label{eq:proper T general}
\Delta P^{\rm T}_{\alpha\beta}
&\equiv  P\,^\rightarrow_{\alpha\beta}[U,W] - P\,^\leftarrow_{\alpha\beta}[U,W]= P\,^\rightarrow_{\alpha\beta}[U,W] - P\,^\rightarrow_{\alpha\beta}[U^*,W]\\
&= 2\bigl(B_{\alpha\beta}\sin\delta + D_{\alpha\beta}\sin 2\delta\bigr) \\
&=  P\,^\rightarrow_{\alpha\beta}[U,W] - \Pbarhat\,^\rightarrow_{\alpha\beta}[U,W] \equiv \Delta P^{\rm CP}_{\alpha\beta}.
\end{aligned}
\end{equation}
The terms $B_{\alpha\beta}$ and $D_{\alpha\beta}$ are the \underline{intrinsic}
carriers of proper T/CP violation. In other words, these terms are odd under $U \to U^*$, as they are proportional to $\sin \delta$. It is transparent that proper T violation and proper CP violation are purely intrinsic and vanish in any matter profile if $U$ is real after rephasing, i.e., $\delta=0$ or $\pi$ using the PMNS matrix of eq.~\eqref{eq:UPDG}. \\

In turn, the $\tau$ asymmetry (improper T) is contaminated with \underline{extrinsic} contributions:
\begin{align}
\label{eq:improper T expanded}
\Delta P^{\tau}_{\alpha\beta}
&= (A_{\alpha\beta}-A_{\beta\alpha})
 +(B_{\alpha\beta}-B_{\beta\alpha})\sin\delta
 +(C_{\alpha\beta}-C_{\beta\alpha})\cos\delta \notag\\
&\quad+(D_{\alpha\beta}-D_{\beta\alpha})\sin 2\delta
 +(E_{\alpha\beta}-E_{\beta\alpha})\cos 2\delta,
\end{align}
since for asymmetric profiles $X_{\alpha\beta} \neq  X_{\beta\alpha}$. Clearly, T and $\tau$ are different for asymmetric matter profiles. \\

For symmetric matter profiles (i.e. $W_{\rm rev} = W$), the coefficients $A, C, E$ are even under the exchange $\alpha \leftrightarrow \beta$, whereas $B$ is odd and D is zero.
Then, 
\begin{align}
\left.\Delta P^{\tau}_{\alpha\beta}\right|_{\mathrm{sym}}
&= 2B_{\alpha\beta}\sin\delta \notag\\
&= \left.\Delta P^{\mathrm{T}}_{\alpha\beta}\right|_{\mathrm{sym}},
\end{align}
and $B_{\alpha \alpha}=0$. Thus, the improper and the proper T tests coincide for a symmetric matter profile. As pointed out in the Introduction, this contributes to the confusion. \\

More importantly, for asymmetric matter profiles and for disappearance probabilities ($\alpha=\beta$) every term vanishes identically in the improper test: $\Delta P^\tau_{\alpha\alpha}\equiv 0$, by definition. On the other hand, the proper test yields 
\begin{align}
    \Delta P^{\rm T}_{\alpha\alpha}
= 2(B_{\alpha\alpha}\sin\delta+D_{\alpha\alpha}\sin 2\delta) \, ,
\end{align}
with non-zero  $B_{\alpha\alpha}$ and $D_{\alpha\alpha}$ for $\alpha =(\mu, \tau)$.  
\\

For electron neutrino disappearance, the form of the Wolfenstein potential  and our choice of the PMNS matrix forces: $B_{ee}=D_{ee}=C_{ee}=E_{ee}=0$.
Thus, $\Delta P^{\rm T}_{ee} = 0$ for any matter profile. The electron neutrino disappearance is special because $R_{23}D_3$ commutes with the matter potential\footnote{For three flavors, if there exists a $U_{ij}(\theta,\delta)$ that commutes with $W$, then $W$ must be diagonal and thus real with two equal entries. Then one can parametrize the PMNS matrix such that $U_{ij}$ is on the left hand end with the remainder being real and thus the probability $P_{kk}$, $k$ different from $i$ or $j$, is invariant under $U \to U^*$. This is satisfied for the standard Wolfenstein matter potential, $(i,j)=(2,3)$ and $k=1(e)$.}. \\

For three flavors of neutrinos, three different $\Delta P^{\rm T}$ are needed to construct all 8 non-zero ones for an arbitrary matter profile. For example, one could choose 
\begin{align}
    \Delta P^{\rm T}_{e\mu}, ~\Delta P^{\rm T}_{\mu e}, ~\Delta P^{\rm T}_{\mu \mu},
\end{align}
then 
\begin{align}
    \Delta P^{\rm T}_{e\tau}=-\Delta P^{\rm T}_{e\mu}, &\quad \Delta P^{\rm T}_{\tau e}
    =-\Delta P^{\rm T}_{\mu e}, \notag \\
    \Delta P^{\rm T}_{\mu \tau} =-\Delta P^{\rm T}_{\mu \mu}-\Delta P^{\rm T}_{\mu e}
    , &\quad 
    \Delta P^{\rm T}_{\tau \mu} =-\Delta P^{\rm T}_{\mu \mu}-\Delta P^{\rm T}_{e\mu}, \notag \\
    \Delta P^{\rm T}_{\tau \tau} 
    =\Delta P^{\rm T}_{\mu \mu}&+\Delta P^{\rm T}_{\mu e} +\Delta P^{\rm T}_{e \mu },
    \label{eq:Tasym}
\end{align}
since $\Delta P^{\rm T}_{ee}=0$. If the matter potential was more general than that given by Wolfenstein, you would need to add a non-zero $\Delta P^{\rm T}_{ee}$ to the above. For three flavors and the form of the matter potential, the PMNS matrix can be factorized into two, a left matrix which contains the complex phase, $R_{23} D_3$, which commutes with the matter potential and a right matrix, $R_{13}R_{12}$, that is real.  In general for $n$-flavor neutrinos, the number of  independent $\Delta P^{\rm T}$'s required is $(n-1)^2$, as the the above $n=3$ special requirements for $\Delta P^{\rm T}_{ee}$ to be zero cannot be satisfied\footnote{Reference \cite{Kuo:1987km} claims the number of independent $\Delta P^{\rm T}$'s is $(n-1)^2-1$ which is not correct for $n>3$, as noted in Ref.~\cite{Akhmedov:2001kd}.}.    The same number  of $\Delta \overline{P}^{\, \rm T}$ is required for anti-neutrinos propagating in matter. This same discussion applies to the proper CP test, since it is equal to the T test. \\

For three flavors and symmetric matter profiles the number of independent
$\Delta P^{\rm T}$ collapse to one, as $\Delta P^{\rm T}_{\alpha \beta}=-\Delta P^{\rm T}_{\beta \alpha}$ which implies that $\Delta P^{\rm T}_{\alpha \alpha}=0$ for all $\alpha$.
Therefore 
\begin{align}
\Delta P^{\rm T}_{e\mu}=\Delta P^{\rm T}_{\mu \tau}=\Delta P^{\rm T}_{\tau e},
\label{eq:Tsym}
\end{align}
which has been discussed in the literature many times, see \cite{Krastev:1988yu, Akhmedov:2001kd, Akhmedov:2004ve}. We emphasize this
simplification is ONLY valid for symmetric matter profiles, where proper and improper T are identical. It is simple to show that eq.~\eqref{eq:Tasym} reduces to eq.~\eqref{eq:Tsym}, using the fact that for symmetric profiles $\Delta P^{\rm T}_{\alpha \beta}$ is odd under interchange of $\alpha \leftrightarrow \beta$. \\

Analogously, for antineutrinos propagating in matter,
\begin{equation}
\begin{aligned}
\label{eq:Pbar_alpha beta}
\Pbar\,^\rightarrow_{\alpha\beta}[U,W] 
&= \overline{A}_{\alpha\beta}
- \overline{B}_{\alpha\beta}\sin\delta
+ \overline{C}_{\alpha\beta}\cos\delta
- \overline{D}_{\alpha\beta}\sin 2\delta
+ \overline{E}_{\alpha\beta}\cos 2\delta \\
&= P\,^\rightarrow_{\alpha\beta}[U^*,-W],
\end{aligned}
\end{equation}
note the definition of the coefficients $\overline{X}_{\alpha \beta} = X_{\alpha \beta}(-W)$, which are not trivially related to $X_{\alpha \beta}$, and are equal only in vacuum. Again, using eq.~\eqref{eq:Ps}  we can easily construct the probability for neutrinos in antimatter as,
$\Phat\,^\rightarrow_{\alpha\beta}[U,W]
=P\,^\rightarrow_{\alpha\beta}[U,-W] $, which changes the sign of $\overline{B}$ and $\overline{D}$ in eq.~\eqref{eq:Pbar_alpha beta}. \\

In the case of the improper cp test,
\begin{align}
\label{eq:improper cp expanded}
\Delta P^{\rm cp}_{\alpha\beta}
&= (A_{\alpha\beta}-\overline{A}_{\alpha\beta})
 +(B_{\alpha\beta}+\overline{B}_{\alpha\beta})\sin\delta
 +(C_{\alpha\beta}-\overline{C}_{\alpha\beta})\cos\delta \notag\\
&\quad+(D_{\alpha\beta}+\overline{D}_{\alpha\beta})\sin 2\delta
 +(E_{\alpha\beta}-\overline{E}_{\alpha\beta})\cos 2\delta,
\end{align}
it already contains extrinsic terms in the presence of a propagation medium. In vacuum, it reduces to the proper CP test, $\Delta P^{\rm CP}_{\alpha\beta} = 2B_{\alpha\beta}\sin\delta$, as the coefficients $\overline{X}_{\alpha \beta} = X_{\alpha \beta}$ and $D$ is zero.
The coefficient $D$ vanishes whenever $W_{\rm rev}=W$ (vacuum and
symmetric matter) and is generically non-zero for asymmetric matter,
except for those channels involving an electron neutrinos,
$D_{e\alpha}=D_{\alpha e}=0$ for all $\alpha$ and any matter profile.\\

\begin{table}[h]
\centering
\renewcommand{\arraystretch}{1.6}
\begin{tabular}{ccccc}
\hline\hline
\text{Proper Test}
& \text{Channel}
& \text{Vacuum }
& \text{Symmetric }
& \text{Asymmetric }\\
& 
& ($W=0$)
& ($W_{\rm rev}=W$)
& ($W_{\rm rev}{\neq}W$)\\
\hline
$\Delta P^{\rm{T}}_{\alpha\beta}
=\Delta P^{\rm CP}_{\alpha\beta}$
& $\alpha\neq\beta$
& $2B_{\alpha \beta}\sin\delta$
& $2B_{\alpha \beta}\sin\delta$
& $2(B_{\alpha \beta}\sin\delta+D_{\alpha \beta}\sin2\delta)$ \\[4pt]
$\Delta P^{\rm{T}}_{\alpha\alpha}
=\Delta P^{\rm CP}_{\alpha\alpha}$
& $\alpha\neq e$
& $0$
& $0$
& $2(B_{\alpha\alpha}\sin\delta+D_{\alpha\alpha}\sin2\delta)$\\[4pt]
$\Delta P^{\rm{T}}_{\alpha\alpha}
=\Delta P^{\rm CP}_{\alpha\alpha}$
%$\Delta P^{\rm{T}}_{ee}$
%$=\Delta P^{\rm CP}_{ee}$
& $\alpha=e$
& $0$ & $0$ & $0$\\[4pt]
\hline
$\Delta P^{\rm CPT}_{\alpha\beta}$
& \text{all}
& $0$ & $0$ & $0$ \\
\hline\hline \nonumber
\end{tabular}
\caption{Three-flavor probability differences for all appearance and disappearance channels and propagation mediums (vacuum, symmetric and  asymmetric matter). $D_{\alpha \beta}=0$ if  $\alpha$ or $\beta$ is $e$-flavor. }
\label{tab:general}
\end{table}

Table~\ref{tab:general} summarises proper T, CP and CPT for all
channels and all matter regimes. An extension of this table, including improper tests as well, is presented in Appendix~\ref{sec:cheat}.

%====================================================================
\section{Two constant-density layers: an example}
\label{sec:2 layers}
%====================================================================
Proper and improper T tests differ for any asymmetric matter profile. The minimal example of such matter profile consists of two consecutive layers, $(a,b)$, with different constant densities, $\rho_a\neq\rho_b$, and propagation lengths $L_a$, $L_b$. For a single layer of density $\rho_k$, the evolution matrix in the rotated basis presented in eq.~\eqref{eq:rotation} is given by\footnote{Throughout this paper we assume a fixed electron fraction $Y_e = 0.5$, which correspond to Earth's crust, in order to convert the electron/positron density in the Wolfenstein potential to matter density. For the standard oscillation parameters, we will use: $\Delta m^2_{21} =7.5 \times 10^{-5}~\mathrm{eV}^2$, $\sin^2\theta_{12} =0.31$, $\Delta m^2_{31} = 2.5 \times 10^{-3}~\mathrm{eV}^2$, $\sin^2\theta_{13} = 0.022$, $\sin^2\theta_{23} = 0.47$.  We assume normal ordering (NO) unless otherwise stated.}
\begin{align}
\widetilde{S}^k_{\alpha\beta}
&= \delta_{\alpha\beta}
+ \widetilde{V}^k_{\alpha 3}\widetilde{V}^k_{\beta 3}(e^{-i 2\phi^k_{31}}-1)
+ \widetilde{V}^k_{\alpha 2}\widetilde{V}^k_{\beta 2}(e^{-i2 \phi^k_{21}}-1),
%{\cal Z}^k_{j1} &= (e^{-i 2\phi^k_{j1}}-1)= 2i e^{-i\phi^k_{j1}} \sin \phi^k_{j1} 
\end{align}
where $\widetilde{V}^k$'s are the  effective mixing matrices in matter which are real in our rotated flavor basis. The kinematic phases controlling the oscillations are
$\phi^k_{j1}=\Delta\lambda_{j1}L_k/(4E_\nu)$, where $\Delta \lambda_{j1}$ are the effective mass splittings in matter.  The $\widetilde{V}^k$'s and the $\Delta \lambda_{j1}$ are easily calculated using the methods in \cite{Abdullahi:2022fkh}. \\

The oscillation lengths are calculated as
\begin{equation}
    L^{\rm osc}_{j1} = \frac{4\pi E_\nu}{\Delta \lambda_{j1}}.
\end{equation}
Note that both $\widetilde{V}^k$ and $\Delta \lambda^k_{j1}$ depend on the product $\rho_k E_\nu$, whereas the oscillation phases, $\phi^k_{j1}$, also depend on the ratio $L_k / E_\nu$. The total evolution matrix for the two layer case is then given by $\widetilde{S}=\widetilde{S}_b\widetilde{S}_a$ (neutrinos traverse layer~$a$ first, then
layer~$b$), and $\widetilde{S}$ depends on the four parameter combinations: $\{\rho_a E_\nu,  L_a/E_\nu\},~~ \{ \rho_bE_\nu, L_b/E_\nu\}$. \\

The muon survival probability, $P\,^\rightarrow_{\mu \mu}$, following eq.~\eqref{eq:P_alpha beta}, can be written as
\begin{equation}
\label{eq:Pmumu full}
P\,^\rightarrow_{\mu\mu}
= A_{\mu \mu}
+ B_{\mu \mu}\sin\delta
+ C_{\mu \mu}\cos\delta
+ D_{\mu \mu}\sin 2\delta
+ E_{\mu \mu}\cos 2\delta.
\end{equation}
The coefficients depend only on the 23 subblock of the evolution matrix $\widetilde{S}$ (see Appendix~\ref{app:coefficients}):
% $\widetilde{S}_{22}$, $\widetilde{S}_{33}$, $\widetilde{S}_{23}$, $\widetilde{S}_{32}$:
\begin{equation}
\label{eq:Pmumu coefficients}
\begin{aligned}
A_{\mu \mu} &= \bigl|c^2_{23}\widetilde{S}_{22} + s^2_{23}\widetilde{S}_{33}\bigr|^2
             + s^2_{23}c^2_{23}\bigl(|\widetilde{S}_{23}|^2 + |\widetilde{S}_{32}|^2\bigr), \\
B_{\mu \mu} &= 2c_{23}s_{23}\,\mathrm{Im}
\!\left[\bigl(c^2_{23}\widetilde{S}_{22}+s^2_{23}\widetilde{S}_{33}\bigr)
\bigl(\widetilde{S}_{32}-\widetilde{S}_{23}\bigr)^*\right], \\
C_{\mu \mu} &= 2c_{23}s_{23}\,\mathrm{Re}
             \!\left[\bigl(c^2_{23}\widetilde{S}_{22}+s^2_{23}\widetilde{S}_{33}\bigr)
             \bigl(\widetilde{S}_{32}+\widetilde{S}_{23}\bigr)^*\right], \\
D_{\mu \mu} &= 2c^2_{23}s^2_{23}\,\mathrm{Im}\bigl[\widetilde{S}_{23}\widetilde{S}^*_{32}\bigr], %\\
\quad E_{\mu \mu} = 2c^2_{23}s^2_{23}\,\mathrm{Re}\bigl[\widetilde{S}_{23}\widetilde{S}^*_{32}\bigr].
\end{aligned}
\end{equation}
For the rest of this paper, we will often drop the subscript $\mu \mu$ for simplicity, since we will focus on $\nu_{\mu}$ disappearance. If there are no subscripts on these coefficients, assume that they are $\mu \mu$.
The $\delta$-odd coefficients, $B$ and $D$, are generically non-zero for $\rho_a\neq\rho_b$ as $\widetilde{S}_{23}\neq\widetilde{S}_{32}$.
When $\rho_a=\rho_b$, there is no asymmetry between layers, and then $B=D=0$, since $\widetilde{S}$ is symmetric.
We are interested in maximizing these $\delta$-odd terms. After checking that $D$ is subleading in general, and for a given pair of $\{\rho_aE_\nu,\rho_bE_\nu\}$, we consider the maximum of $\abs{B}$ over any pair of propagation lengths $\{L_a,L_b\}$. The case $\rho E_\nu < 0$ corresponds to antineutrinos propagating in matter, that is, it is related to $\overline{X}_{\alpha \beta}$ in eq.~\eqref{eq:Pbar_alpha beta}.\\

\begin{figure}
    \centering
    \includegraphics[width=\linewidth]{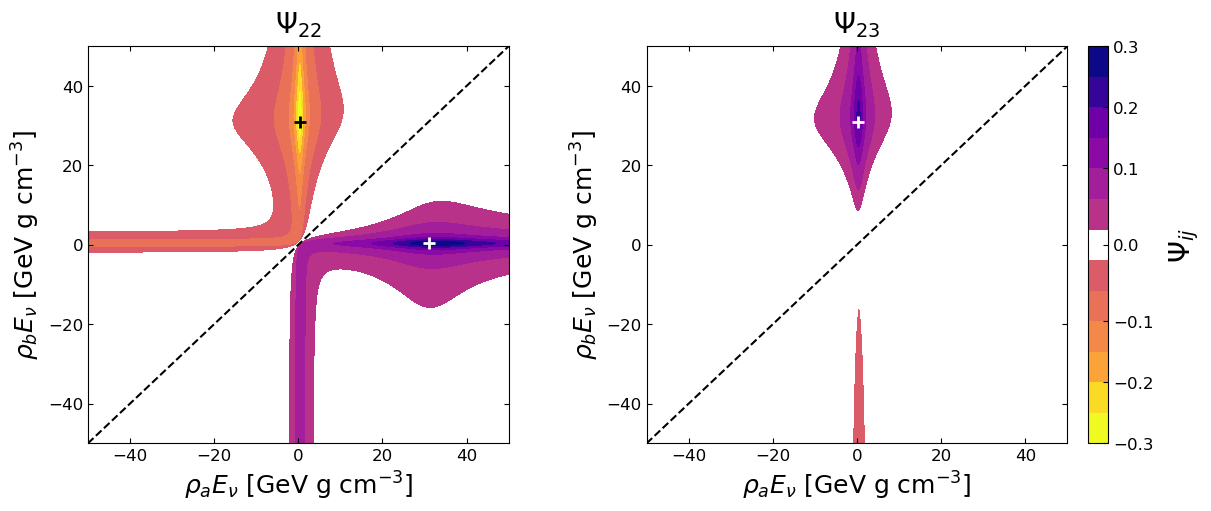}
    \caption{Terms that multiply the oscillation factors in the asymmetry $\widetilde{S}_{32} - \widetilde{S}_{23}$, which determines the size of the leading T/CP-violating contribution in $P_{\mu \mu}$, as a function of the product $\rho E_{\nu}$ for each one of the two layers of constant density. 
    The peaks of the terms are found close to the solar and atmospheric resonances, marked by plus signs.
    When $\rho_a = \rho_b$, along the dashed black line, the profile is symmetric and $\widetilde{S}_{32} = \widetilde{S}_{23}$, thus all terms must be exactly zero.}
    \label{fig:psi_ij}
\end{figure}

For the kinematic phases we use
\begin{align}
{\cal Z}^k_{j1} & \equiv  (e^{-i 2\phi^k_{j1}}-1)= -2i e^{-i\phi^k_{j1}} \sin \phi^k_{j1},
\end{align}
then the relevant difference for $B$ is given by combinations of the elements of $\widetilde{V}$
\begin{align}
    \widetilde{S}_{32}-\widetilde{S}_{23} &=\sum_\beta (\widetilde{S}^b_{3\beta} \widetilde{S}^a_{\beta 2} -\widetilde{S}^b_{2 \beta} \widetilde{S}^a_{\beta 3} ) \notag \\
    &= {\cal Z}^b_{31} {\cal Z}^a_{31} ~(\sum_\beta \widetilde{V}^b_{\beta 3}\widetilde{V}^a_{\beta 3}) ~[\widetilde{V}^b_{33} \widetilde{V}^a_{23}-\widetilde{V}^a_{33} \widetilde{V}^b_{23}]  \notag \\
    &+ \,{\cal Z}^b_{31} {\cal Z}^a_{21} ~[\sum_\beta \widetilde{V}^b_{\beta 3}\widetilde{V}^a_{\beta 2}] ~(\widetilde{V}^b_{33} \widetilde{V}^a_{22}-\widetilde{V}^a_{32} \widetilde{V}^b_{23}) \notag \\
    &+\, {\cal Z}^b_{21} {\cal Z}^a_{31} ~[\sum_\beta \widetilde{V}^b_{\beta 2}\widetilde{V}^a_{\beta 3}] ~(\widetilde{V}^b_{32} \widetilde{V}^a_{23}-\widetilde{V}^a_{33} \widetilde{V}^b_{22})  \notag \\
    &+ \,{\cal Z}^b_{21} {\cal Z}^a_{21} ~ (\sum_\beta \widetilde{V}^b_{\beta 2}\widetilde{V}^a_{\beta 2}) ~[\widetilde{V}^b_{32} \widetilde{V}^a_{22}-\widetilde{V}^a_{32} \widetilde{V}^b_{22}] \notag \\
    &\equiv \sum_{i, j}  {\cal Z}^b_{j1} {\cal Z}^a_{i1} \Psi_{ij} \,.
    \label{eq:asymmetry}
\end{align}
In the last row we have defined the coefficients that multiply the oscillation terms, $\Psi_{ij} =  (\sum_\beta \widetilde{V}^b_{\beta j}\widetilde{V}^a_{\beta i}) ~(\widetilde{V}^b_{3j} \widetilde{V}^a_{2i}-\widetilde{V}^a_{3i} \widetilde{V}^b_{2j}) $.
If the density of the two layers are the same, all of these 4 terms vanish as the terms $[\cdots] =0$ in eq.~\eqref{eq:asymmetry}. The coefficient $\Psi_{22}$, which multiplies the ${\cal Z}^b_{21} {\cal Z}^a_{21}$ term, has two broad peaks, one when $\rho_a E_\nu$ is near the solar resonance and $\rho_b E_\nu$ is near the atmospheric resonance AND a second peak when the resonances are interchanged, as shown in the left panel of Fig.~\ref{fig:psi_ij}. These peaks are approximately $\pm 0.3$. 
In the right panel of Fig.~\ref{fig:psi_ij}, we show $\Psi_{23}$, which has one broad peak which coincides with one of the two peaks of $\Psi_{22}$. Similarly, $\Psi_{32}$, which is not shown, coincides with the other peak of $\Psi_{22}$. 
$\Psi_{33}$ is never significant because for any $\rho E$, $|\widetilde{V}_{23}(\rho E_\nu)| \ll 1$.\\

Choosing $\rho_a E_\nu$ to be near the solar resonance, then $|{\cal Z}^a_{21}|$ is maximized if the path length, $L_a$, is on the order of 15,000 km/GeV.  With $\rho_b E_\nu$ near the atmospheric resonance, 
then both $|{\cal Z}^b_{21}|$ and $|{\cal Z}^b_{31}|$ are maximized at path lengths of order 500 - 1000 km/GeV. Therefore we expect the T-violating asymmetry for $\nu_\mu$ disappearance to be large in a scenario when one layer is near the solar resonance with a path length approximately the solar oscillation length and the other layer is near the atmospheric resonance and with a path approximately the atmospheric oscillation length.\\

\begin{figure}
    \centering
    \includegraphics[width=\linewidth]{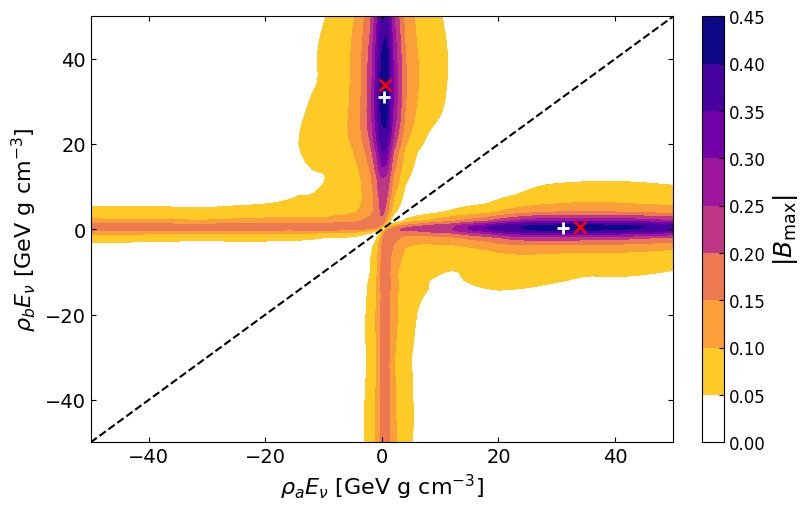}
    \caption{Maximum size of the $\sin \delta$ term in $P_{\mu \mu}$, the coefficient $B$, as a function of $\rho E_\nu$ in two layers of constant density, after maximizing it over any pair of propagation lengths, $L_a$ and $L_b$.
    The effect is larger when one of the layers is close to the solar resonance ($\rho E_\nu \sim 0.34 ~\rm{GeV \, g \, cm^{-3}}$) and the other one is close to the atmospheric resonance ($\rho E_\nu \sim 31 ~\rm{GeV\, g\, cm^{-3}}$), marked by white plus signs.
    In particular, the global maximum is $\abs{B} \sim 0.4$ (red crosses) found for $\rho_a E_\nu \approx 0.7~\rm{GeV \, g \, cm^{-3}}$, $\rho_b E_\nu \approx 34~\rm{GeV \, g \, cm^{-3}}$. For an energy of $1$~GeV, the path lengths that yield the maximum $B$ correspond to $L_a \approx 21\,000$~km and $L_b \approx 830$~km.
     When $\rho_a = \rho_b$, along the dashed black line, the profile is symmetric and $B$ must be exactly zero.  }
    \label{fig:Bmax}
\end{figure}

In Fig.~\ref{fig:Bmax} we show that the maximum size of $B$ can be as large as $\abs{B} \sim 0.4$, and thus compete with the leading term of the probability, $A$. In particular, the maximum value of $\abs{B}$ (red crosses) is found for one of the layers with $\rho E_\nu \approx 0.7~\rm{GeV \, g \, cm^{-3}}$, and the other with $\rho E_\nu \approx 34~\rm{GeV \, g \, cm^{-3}}$. We emphasize that this result is close to one layer being on the solar resonance ($\rho E_\nu \sim 0.34 ~\rm{GeV \, g \, cm^{-3}}$) and the other on the atmospheric resonance ($\rho E_\nu \sim 31 ~\rm{GeV\, g\, cm^{-3}}$), marked by white plus signs. This implies that this effect would be easier to measure with neutrino probabilities, $P_{\mu \mu}$, rather than with $\Pbar_{\mu \mu}$, since we are assuming normal ordering. In turn, for inverted ordering the resonances are found for antineutrinos, and the results are similar to a rotated version of Fig.~\ref{fig:Bmax}, in which $E \to -E$. $B$ is sensitive to the ordering of the layers, and the exchange $a \leftrightarrow b$ leads to a change of sign\footnote{This is only true for disappearance channels. See Appendix~\ref{app:Cmax} for more details.}, as it could have been anticipated by eq.~\eqref{eq:asymmetry}. This is a consequence of the reversal identity, eq.~\eqref{eq:master}, since backwards propagation flips the sign of the T/CP-violating terms. \\

\begin{figure}
    \centering
    \includegraphics[width=0.7\linewidth]{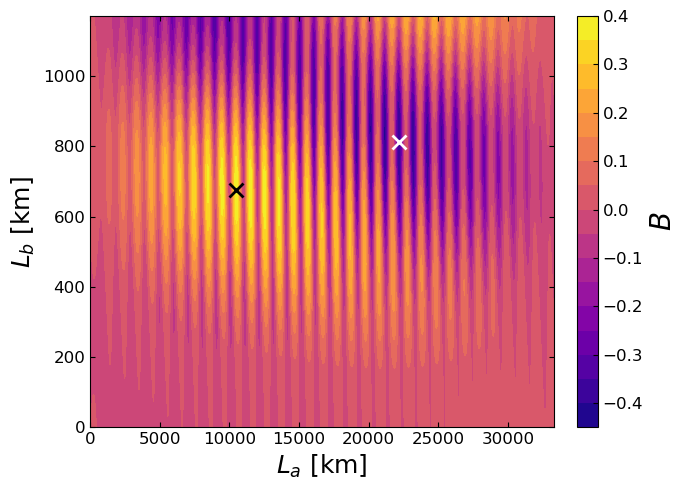}
    \caption{Size of $B$ in $P_{\mu \mu}$ as a function of the propagation lengths $\{ L_a, L_b \}$ for the configuration that maximizes $B$: $\rho_a E_\nu \approx 0.7~\rm{GeV \, g \, cm^{-3}}$, $\rho_b E_\nu\approx 34~\rm{GeV \, g \, cm^{-3}}$, close to the solar and atmospheric resonance values. The energy has chosen to be $E_\nu =1~\rm{GeV}$. $L_a$ must be close to the oscillation length $L_{21}^{\rm osc}$ to make $B$  sizable.  The crosses indicate the location of the maximum values of $|B|$ for each of the two regions. The values of the path lengths at these two maximums are given by $(L_a, L_b) \,\approx \,(10\,500,~700)$~km and $(21\, 000,~830)$~km.}
    \label{fig:BvsLaLb}
\end{figure}

Note that for fixed $\{\rho_a E_\nu, \rho_bE_\nu\}$, the oscillation parameters $\{L_a/E_\nu, L_b/E_\nu\}$ just control the length scale at which we could find the maximum effect. The layer near the solar resonance must have a propagation length close to half the oscillation length associated to $\Delta \lambda_{21} \approx \Delta m^2_{21}$ (since $\rho E_\nu$ is small), and this leads to $L/E_\nu \sim 15000~\mathrm{km\, GeV^{-1}}$. For GeV neutrinos, this implies large baselines, as we show in Fig.~\ref{fig:BvsLaLb}, where we have fixed $E_\nu = 1$~GeV. The products $\rho_a E_\nu \approx 0.7~\rm{GeV \, g \, cm^{-3}}$, $\rho_b E_\nu \approx 34~\rm{GeV \, g \, cm^{-3}}$ correspond to the maximum value found for $B$ (red crosses in Fig.~\ref{fig:Bmax}). The required propagation length, $L_a$, is larger than $L_b$, since we chose $\rho_a < \rho_b$. These required baselines may be reduced decreasing the energy, but to obtain the same size for $B$, the density must be accordingly increased to conserve the product $\rho E_\nu$. \\

\begin{figure}
    \centering
    \includegraphics[width=0.7\linewidth]{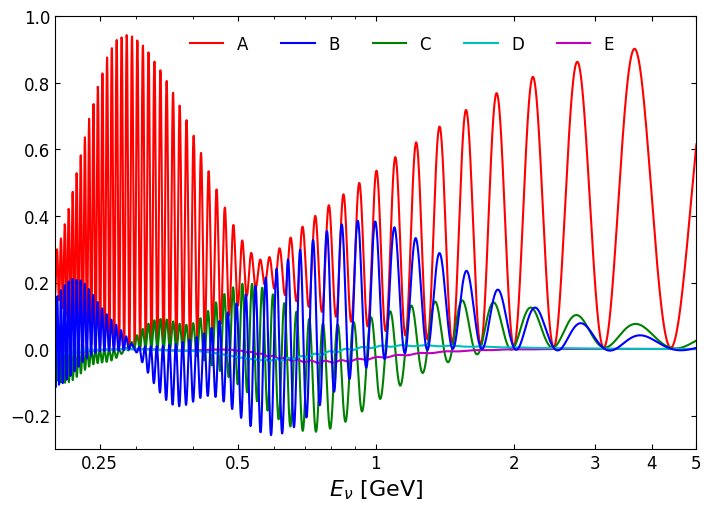}
    \caption{Muon survival probability for a idealistic scenario optimized to measure $\sin \delta$ at $\sim 1$~GeV: two layers with $\rho_a E_\nu \approx 0.7~\mathrm{g \, cm^{-3}}$, $\rho_b E_\nu \approx 34~\mathrm{g \, cm^{-3}}$, $L_a \approx 10500$~km, $L_b \approx 700$~km. The $\sin \delta$ coefficient ($B$) is large in the GeV range and it could be measured.}
    \label{fig:ideal terms}
\end{figure}

Focusing on the maximum possible effect, but restricting to baselines as small as possible, we will take the maximum value of $B$ in the yellow region, which corresponds to the left cross in Fig.~\ref{fig:BvsLaLb}, where we have chosen the neutrino energy to be $E_\nu = 1$~GeV. This would be an ideal, unrealistic experiment composed of two layers with density $\rho_a \approx 0.7~\rm{g \, cm^{-3}}$, $\rho_b \approx 34~\rm{g \, cm^{-3}}$ and propagation lengths $L_a \approx 10\,500$~km, $L_b \approx 700$~km. After fixing the propagation lengths, we study the effect as a function of the neutrino energy. Such experiment would be optimized to measure T/CP violation through $P_{\mu \mu}$ using GeV neutrinos, as shown in Fig.~\ref{fig:ideal terms}, where we plot the terms of the probability defined in eqs.~\eqref{eq:Pmumu full} \& \eqref{eq:Pmumu coefficients}. Note that the magnitude of $B$ is large, and it could contribute significantly to the probability, depending on the value of $\delta$. 

\begin{figure}
    \centering
    \includegraphics[width=0.87\linewidth]{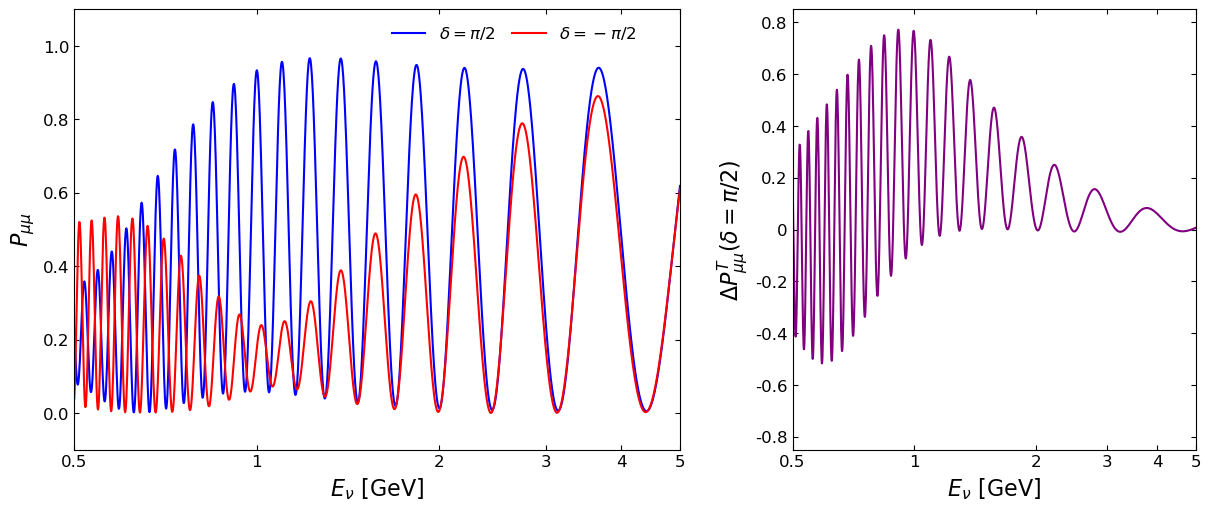}
    \caption{Muon survival probability for the same scenario as in Fig.~\ref{fig:ideal terms}. In the left panel, $P_{\mu \mu}$ for maximal CP violation, with $\delta = \pi/2$ (blue) or $\delta = -\pi/2$ (red). In the right panel, the difference between both probabilities, which amounts to the proper T test. Explicitly, $\Delta P^{\rm T}_{\mu \mu} = 2 B_{\mu \mu} $. }
    \label{fig:P delta}
\end{figure}

The role of the T/CP-odd part is highlighted in Fig.~\ref{fig:P delta}. In the left panel, we show $P_{\mu \mu}$ for $\delta = \pi/2$ (blue line) and for $\delta = -\pi /2$ (red line). Note that the difference between both probabilities, $P^\rightarrow_{\mu \mu} [U, W] - P^\rightarrow_{\mu \mu} [U^*, W]$, is just the proper T test in eq.~\eqref{eq:proper T general}, $\Delta P^{\rm T}_{\mu \mu} = 2(B \sin \delta + D\sin 2\delta)$. We evaluate $\Delta P^{\rm T}_{\mu \mu} $ for $\delta = \pi /2$ in the right panel of Fig.~\ref{fig:P delta}. Because of how we chose the rest of parameters, it peaks around $1$~GeV, with a maximum value of $2B_{\rm max} \sim 0.8$.

For an experiment that is restricted to only low $\rho E$, or its matter profile is prominently symmetric, another possibility to determine the T/CP phase is through measuring the T/CP-conserving part of the disappearance probability, as already proposed in Ref.~\cite{Denton:2023qmd}. This would imply measuring for example the coefficient $C$, which is proportional to $\cos \delta$. Unlike $B$, this term is not dependent on the asymmetry, and thus does not vanish if the matter profile is symmetric. Furthermore, it has the advantage of being relatively large even for low $\rho E_\nu$, see Appendix~\ref{app:Cmax} for more details.

% \clearpage
%====================================================================
\section{Experimental sensitivity to T/CP violation through $P_{\mu\mu}$ in DUNE and HK}
\label{sec:DUNE+HyperK}
%====================================================================
As we already pointed out throughout this paper, it is possible to measure T/CP-violating terms just by measuring $P_{\mu \mu}$, since it is dependent on $\sin \delta$, and one of the reasons this was overlooked until now is the conflation between proper and improper time-reversal tests (see Section~\ref{sec:PandImp}), which differ for asymmetric matter profiles. Note that the proper T test for muon disappearance, $\Delta P^{\rm T}_{\mu \mu}$, would require measuring both the neutrino in forward and backward propagation ($P^\rightarrow_{\mu \mu}$ and $P^\leftarrow_{\mu \mu}$), which is unpractical since experimental detectors and neutrino sources are challenging to interchange\footnote{Note that a proper CP test, which also isolates the $\delta$-odd part, is totally unfeasible since it requires a propagation medium composed of antimatter. In addition, we highlight that comparing $P_{\mu \mu}$ to $\Pbar_{\mu \mu}$ in matter is an improper cp test and mixes intrinsic and extrinsic contributions.}. Even if it is not practically possible to isolate the $\delta$-odd part of the probability by performing a proper T test, one could ask whether current or upcoming experiments are sensitive to the $\delta$-odd terms in $P_{\mu \mu}$. Recall that the decomposition of the muon neutrino survival probability in T/CP-violating terms ($B, D$) and T/CP-conserving terms ($A, C, E$), is written as
\begin{equation}
    P_{\mu \mu} = A_{\mu \mu} + B_{\mu \mu}\sin{\delta} + C_{\mu \mu} \cos{\delta} + D_{\mu \mu}\sin{2\delta} + E_{\mu \mu} \cos{2 \delta}. \nonumber
\end{equation}
We already proved that $B$ can be quite large in some experimental setups in Section~\ref{sec:2 layers}, and now we want to test how relevant $B$ and $D$ are in current or upcoming experiments. \\

\begin{figure}
    \centering
    \includegraphics[width=\linewidth]{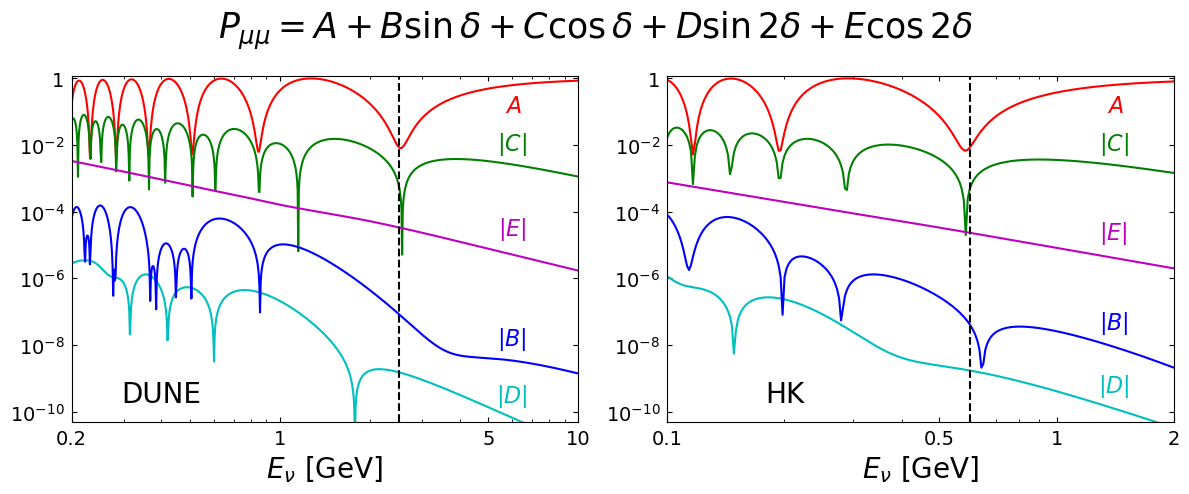}
    \caption{Terms contributing to $P_{\mu\mu}$ for DUNE (left) and HK (right). The vertical dashed line is the neutrino beam peak energy, which is located near the first oscillation minimum. The T/CP-violating terms are $B$ and $D$, which are suppressed with respect to the T/CP-conserving ones. Larger asymmetries and longer propagation lengths are required to enhance the sensitivity to $\sin \delta$.
    These results are for neutrinos, but similar results are obtained for antineutrinos.}
    
    \label{fig:DUNE and T2K}
\end{figure}

The most promising experimental setups are LBL experiments, since long baselines are required to enhance $B$ at the GeV scale.
In Fig.~\ref{fig:DUNE and T2K}, we show the T/CP-dependence of the $\nu_\mu$ disappearance probability, $P_{\mu \mu}$, for DUNE and for HK. In both cases, at the experiment peak neutrino energy, which is tuned to be around the first oscillation minimum ($E_\nu\sim 2.6$~GeV for DUNE, and $E_\nu\sim 0.6$~GeV for HK), the magnitude of $B$ is of the order of $10^{-7}$, being completely negligible. Even more suppressed is the other $\delta$-odd term, $D$. An enhancement of up to 3 orders of magnitude is found for low energies, but it would still be a per mill contribution to the probability.
We can conclude that this effect is not measurable in DUNE or HK. This strong suppression can be understood in terms of the two-layer system considered in Section~\ref{sec:2 layers}: both DUNE and
HK operate at experimental configurations far from the optimal configuration, which required larger asymmetries and propagation lengths. \\

$A$ is always positive, whereas $B, C, D$ oscillate between positive and negative values and their downwards spikes in Fig.~\ref{fig:DUNE and T2K} indicate a change of sign. $E$ does not oscillate in the chosen energy range, and this can be understood through the constant-density layer approximation, in which the oscillations of $E$ are governed by $\mathcal{Z}_{21}$, which sets a large $L/E_\nu$, determined by $\Delta m^2_{21}$.\\

\begin{figure}
    \centering
    \includegraphics[width=0.9\linewidth]{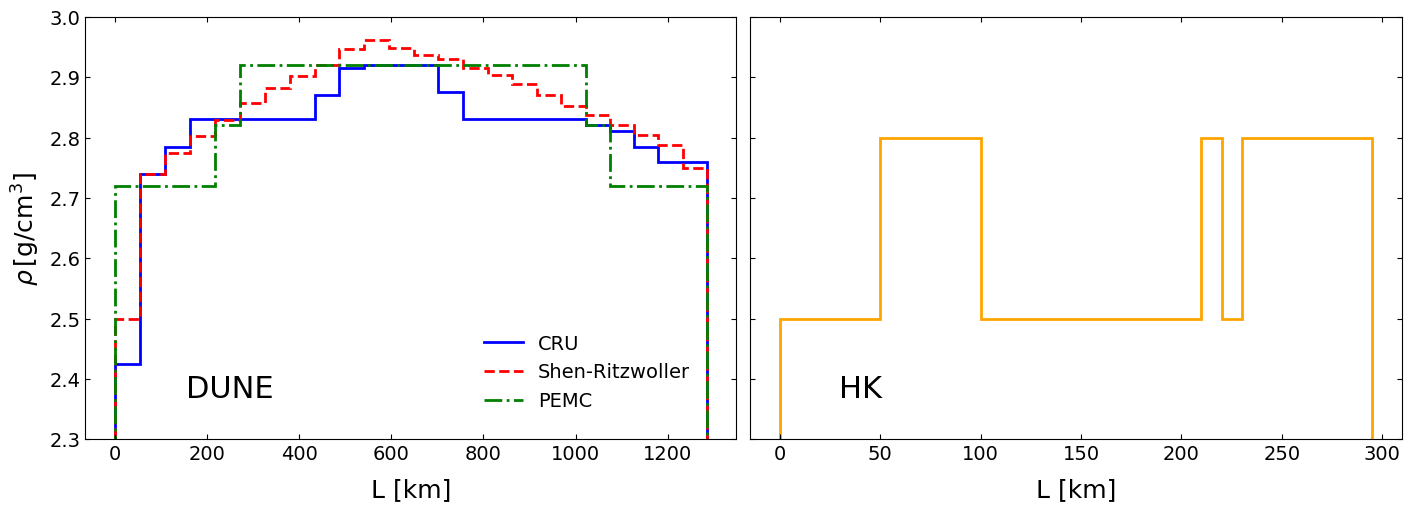}
    \caption{Matter density profiles assumed for DUNE~\cite{Roe:2017zdw} (left panel) and HK~\cite{Hagiwara:2011kw} (right panel). For DUNE, any of the three different profiles lead to minimal changes in the probability, as already shown in Refs.~\cite{Roe:2017zdw, Kelly:2018kmb}. Note that approximating the matter profile to a constant density, which is symmetric, would eliminate the T/CP-violating terms of $P_{\mu \mu}$ in Fig.~\ref{fig:DUNE and T2K}.}
    \label{fig:density_profiles}
\end{figure}

In Fig.~\ref{fig:density_profiles} we present the matter density profiles that we have assumed for DUNE and HK. Note that considering a constant density would completely cancel out the T/CP-violating terms, $B$ and $D$, shown in Fig.~\ref{fig:DUNE and T2K}. In the right panel, the matter profile used for HK, taken from Ref.~\cite{Hagiwara:2011kw}. In the left panel, the three different choices to model the matter density profile that the neutrino beam traverses until reaching the far detector in DUNE: Crustal, Shen-Ritzwoller and PEMC. We checked that the discrepancies between them are of the per mill level on the probabilities, in line with the results found in Ref.~\cite{Roe:2017zdw}. When comparing one of these matter profiles with a single layer with the same average density, the odd terms in $\delta$ (i.e., $B$, $D$) vanish, since the constant density profile is symmetric. Nevertheless, since these contributions are small for DUNE, the changes in $P_{\mu \mu}$ are also at the per mill level. For DUNE, the uncertainty on the density profile shape does not alter significantly the expected oscillations probabilities, being the average density the most important feature~\cite{Kelly:2018kmb}. We highlight that this is the case because the asymmetry in the matter profile is not large enough, but this is not necessarily true for all possible experiments, as we will show in the next section.

%====================================================================
\section{Possible futuristic experiments}
\label{sec:futuristic experiments}
%====================================================================

In this section, we will provide some experimental setups where the T/CP-violating part of $P_{\mu \mu}$ is large ($B \sim 0.2-0.3$) and the rest of physical parameters lead to somewhat feasible, yet futuristic experimental setups. We emphasize that to this end, it is indispensable to consider a very asymmetric matter profile. Also, the path length in the higher density region is set by $\Delta m^2_{31}$, $500$ km/GeV, whereas the path length in the low density region is set by $\Delta m^2_{21}$, $15,000$ km/GeV. It is this latter path length that is difficult to achieve on or near the Earth's surface. We present two scenarios based on the two-layer scheme presented in Section~\ref{sec:2 layers}, which exploit the asymmetry between propagation in vacuum and propagation through the Earth: (1) Earth to geostationary satellite experiment and (2) Moon to Earth experiment. In both scenarios, neutrinos would propagate through vacuum for very large distances, and we will consider that they would roughly traverse the whole Earth diametrically. Since the Earth has an approximate symmetric matter profile, we will treat the problem in the two-layer approximation, with $\rho_a = 0$, $\rho_b = 8~\mathrm{g \, cm^{-3}}$, and then we will compare with a more realistic description of the Earth matter profile, using the PREM model~\cite{Dziewonski:1981xy}, which is symmetric.

\subsection{Earth to geostationary satellite experiment}
This experiment consists of an accelerator neutrino beam of $E_\nu \sim 5$~GeV that traverses through the Earth to a detector in geostationary orbit on the other side of the Earth. The propagation path length through the Earth is set to its diameter, $2R_E=12742$~km. While the propagation from the Earth's surface to the detector, which is approximately vacuum, is set to $3.6 \times 10^{4} $~km, which is roughly the altitude of a geostationary orbit. 
The geostationary orbit is chosen to ensure that the fixed neutrino beam continuously points toward the detector, which appears stationary relative to the Earth's surface. Therefore, the satellite orbit should lie on the equatorial plane. \\

\begin{figure}
\centering
\textbf{\Large Earth to geostationary satellite}\vspace*{1cm}

    \hspace*{0.6cm}\includegraphics[width=0.4\linewidth]{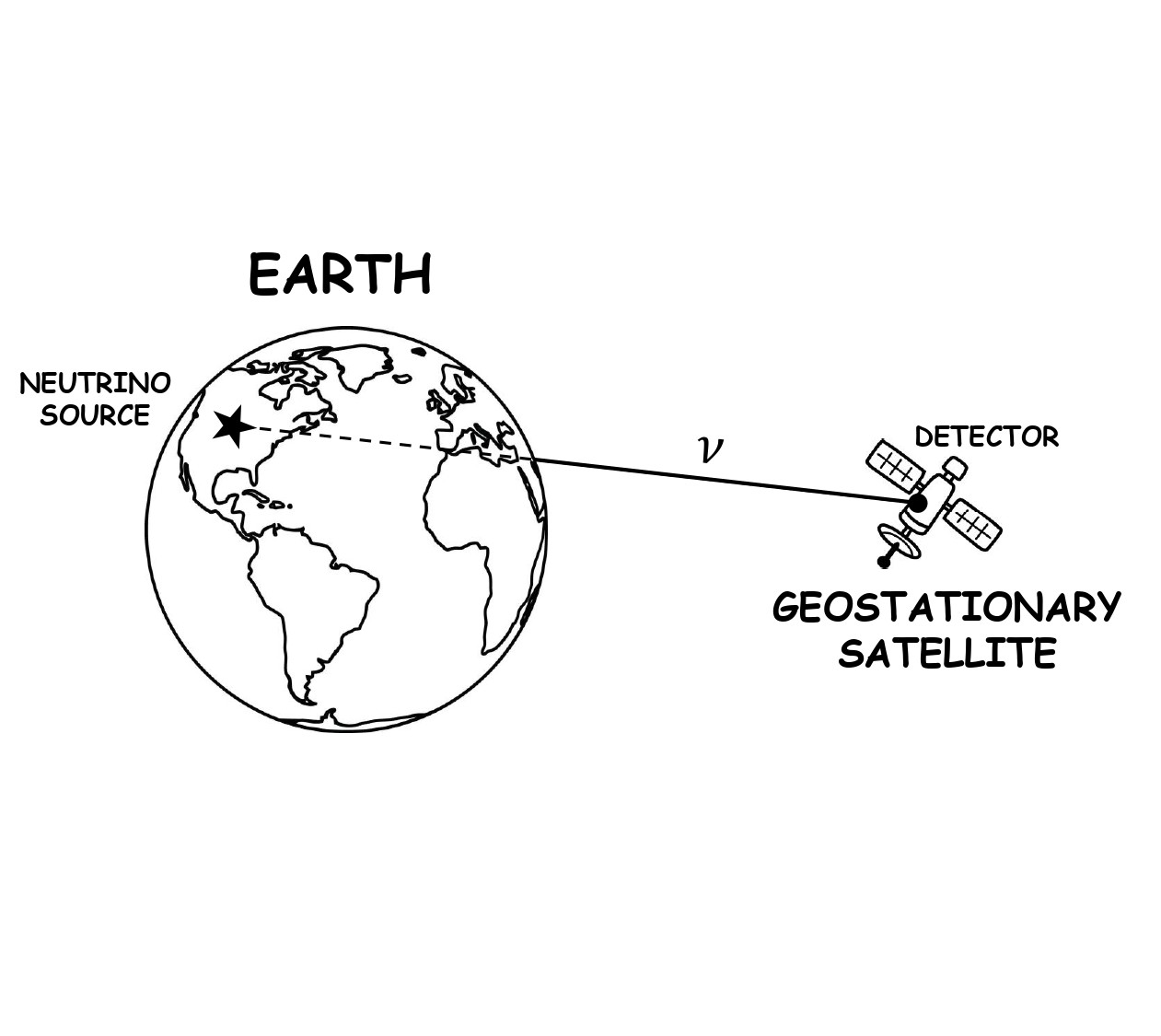}\includegraphics[width=0.49\linewidth]{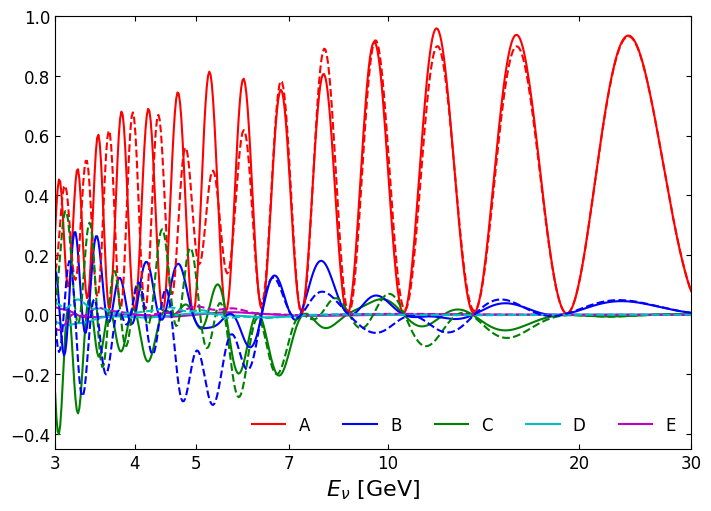}
    \includegraphics[width=0.9\linewidth]{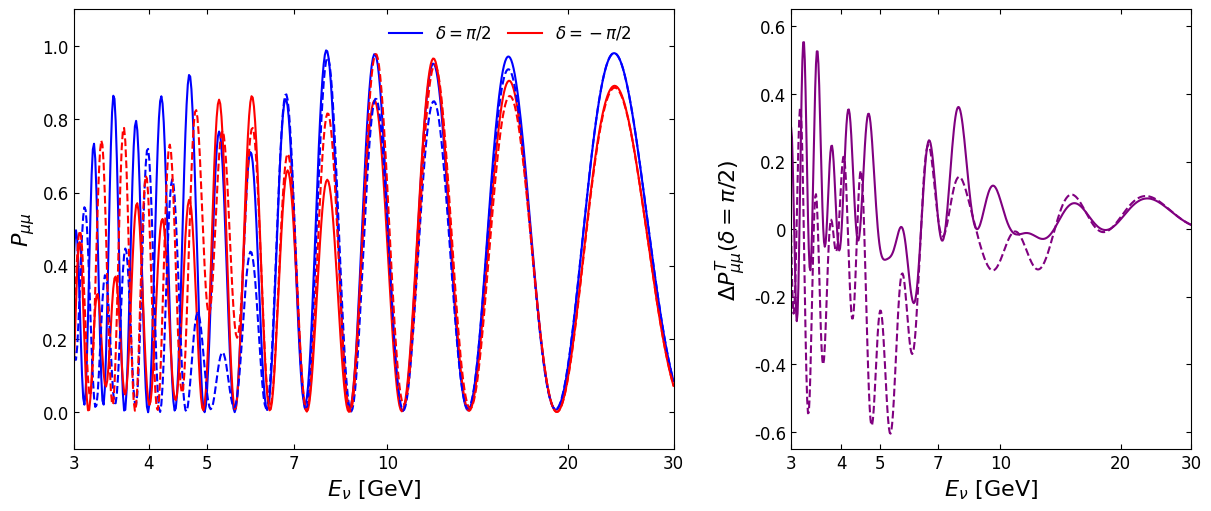}
    \caption{Futuristic experiment in which a neutrino accelerator beam of $E_\nu \sim 5$~GeV passes through the Earth and arrives to a neutrino detector in an equatorial, geostationary orbit (upper left panel). Terms of $P_{\mu \mu}$ as a function of the neutrino energy (upper right panel), showing that the T/CP-violating part ($B$) is large. In the lower panels, $P_{\mu \mu}$ for maximal CP violation, i.e., $\delta = \pm \pi/2$ (left panel), and the difference between them, which is the proper T test, $\Delta P^{\rm T}_{\mu \mu}$ (right panel).
    Assuming the PREM model for the Earth density (dashed lines) differs with respect to approximating the Earth as a constant-density layer (solid lines), specially for $E_{\nu} \sim 5$~GeV, which proves that the experiment would be sensitive to the Earth's inner structure. 
    For this experiment, T/CP-violating terms in $P_{\mu \mu}$ would be as large as $\sim 0.3$.}
    \label{fig:scenario 2}
\end{figure}

In Fig.~\ref{fig:scenario 2}, we show the sketch of the experiment (upper left panel), which would measure sizable T/CP-violating terms in $P_{\mu \mu}$ (upper right panel). Lower left panel shows $P_{\mu \mu}$ for maximal T/CP-violation, i.e. for $\delta = +\pi/2$ (blue line) or $\delta = -\pi/2$ (red line). The difference between both probabilities is given by the proper T test, $\Delta P^{\rm T}_{\mu \mu} (\delta = \pi/2)= 2B$, shown in the lower right panel. Solid lines correspond to the two-layer approximation and dashed lines implement a more realistic PREM profile for the Earth. \\

The large difference between solid and dashed lines, specially for around $5$~GeV, shows that this experiment would be sensitive to the Earth's inner structure. Thus, the asymmetry between the matter density in the mantle and in the core would enhance the T/CP violation in $P_{\mu \mu}$ measured by this experiment\footnote{Note that the propagation length through the Earth depends on the latitude at which the neutrino beam is located. A path length of $2R_E$ assumes that the beam is located in the equator and it is pointing to its antipodes, i.e., that it is deflected $90^\circ$. However, we have checked that for a latitude of $45^\circ$, the neutrino beam would need to be deflected $\sim 50 ^\circ$, and even if the Earth core would not be crossed, the effect would still be sizable. Our conclusions are then valid for a wide range of neutrino beam locations.}. This proves that, unlike DUNE or HyperK, there could be experiments which are highly sensitive to the shape of the matter profile, independently of the average matter density.

\subsection{Moon to Earth experiment}
In this scenario, a neutrino source of $E_\nu \sim30$~GeV which is located on the Moon, points to the Earth, and is detected by multiple terrestrial detectors. The propagation in vacuum is $3.85 \times 10^{5}$~km, which is roughly the Earth-Moon distance and $2R_E$, as in the previous experiment. Note that for this scenario, the larger propagation in vacuum implies a larger neutrino energy. We highlight that from the Moon, the Earth is approximately fixed in the sky. We have ignored the ``lunar night-day asymmetry" produced by the Earth rotation. \\

\begin{figure}
\centering
\textbf{\Large Moon to Earth}\vspace*{1cm}

    \includegraphics[width=0.45\linewidth]{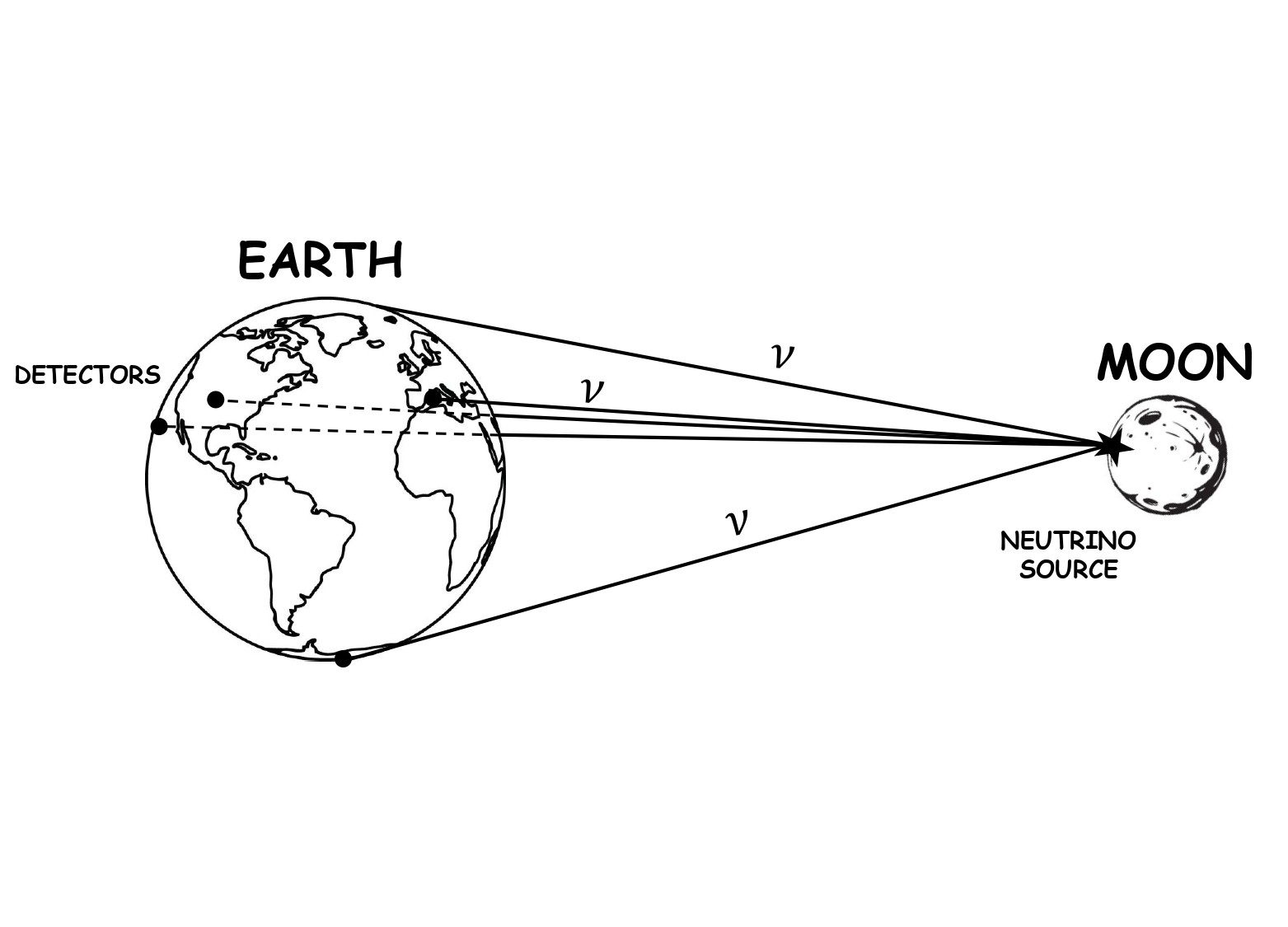}\includegraphics[width=0.49\linewidth]{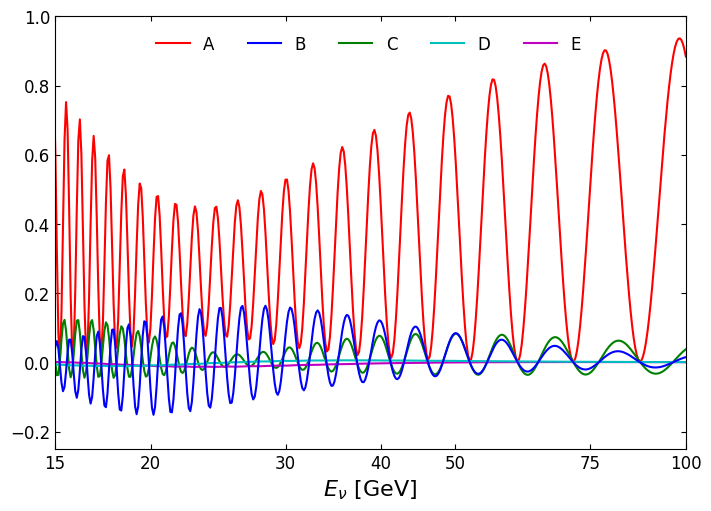}
    \includegraphics[width=\linewidth]{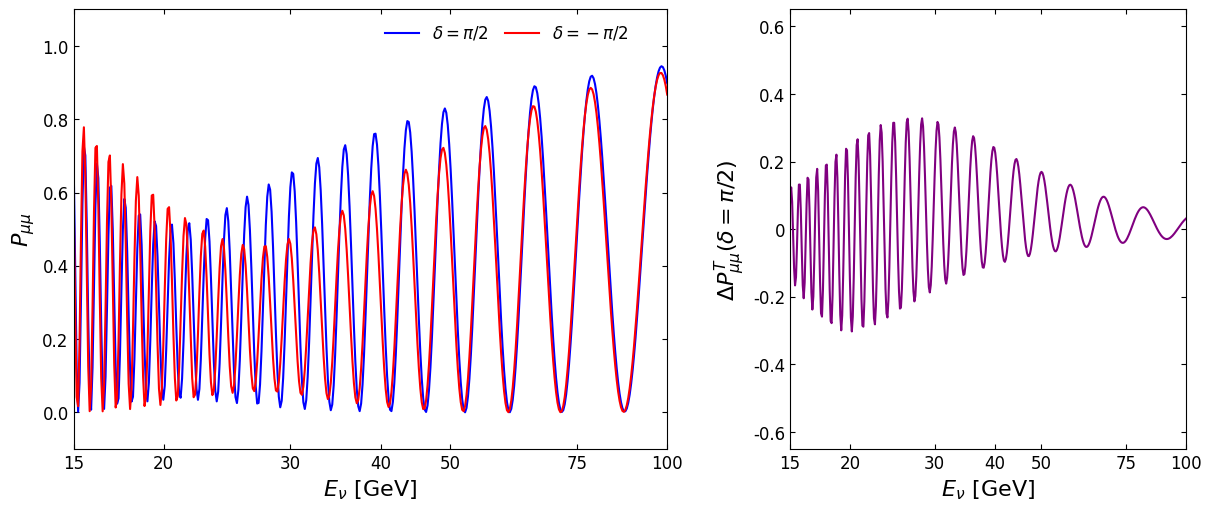}
    \caption{Futuristic experiment in which a neutrino source located in the Moon showers terrestrial detectors with $\sim 30$~GeV neutrinos (upper left panel). Terms of $P_{\mu \mu}$ as a function of the neutrino energy (upper right panel), showing that the T/CP-violating part ($B$) is large. We assume that the Earth is crossed diametrically, and we ignore the night-day asymmetry. In the lower panels, $P_{\mu \mu}$ for maximal CP violation, i.e., $\delta = \pm \pi/2$ (left panel), and the difference between them, which is the proper T test, $\Delta P^{\rm T}_{\mu \mu}$ (right panel). We approximate the Earth as a constant-density layer, since considering the PREM model does not make any difference for this energy range.
    For this experiment, T/CP-violating terms in $P_{\mu \mu}$ would be as large as $\sim 0.15$.}
    \label{fig:scenario 1}
\end{figure}

In Fig.~\ref{fig:scenario 1}, we show that the Moon to Earth experiment (sketched in the upper left panel) would also yield measurable T/CP-violating terms to $P_{\mu \mu}$ (upper right panel). Lower left panel shows $P_{\mu \mu}$ for maximal T/CP-violation, i.e. for $\delta = +\pi/2$ (blue line) or $\delta = -\pi/2$ (red line). The difference between both probabilities is given by the proper T test, $\Delta P^{\rm T}_{\mu \mu} (\delta = \pi/2)= 2B$, shown in the lower right panel.  The solid lines correspond to the two-layer approximation, whereas replacing the constant density of $8~\mathrm{g \, cm^{-3}}$ by the PREM model does not considerably affect the results. This experiment leads to a slightly smaller T/CP-violation in $P_{\mu \mu}$ in comparison with the previous one, but it has the advantage of leveraging all existing terrestrial neutrino detectors. \\

We conclude that both experiments would have a great sensitivity to $\sin \delta$ just by measuring muon neutrino disappearance, $P_{\mu \mu}$.

\clearpage

%====================================================================
\section{Conclusions}
\label{sec:conclusions}
%====================================================================

In this paper we have provided a coherent and comprehensive overview on T, CP and CPT for neutrino propagation in an arbitrary matter profile both for the proper tests and the improper tests. Proper tests change the propagation environment 
as well as the neutrino states, whereas the improper tests only change the neutrino states, that is, they only affect the vacuum part of the Hamiltonian. For the proper CP test this means that not only do neutrinos get changed to anti-neutrinos but the matter has to be changed to anti-matter. Only in vacuum do the proper and improper CP tests coincide.
Whereas for proper T-violation not only the initial and final flavors need to be interchanged but one has to reverse the direction of propagation. Only in symmetric matter profiles do the proper and improper T tests coincide as reversing the matter profile gives the same profile.
Similarly for the proper and improper CPT test, which is a concatenation of CP and T. The proper and improper  CPT tests only coincide in vacuum.\\

In this paper we give a detail proof that the proper T test in the $\nu_\mu$ and $\nu_\tau$ disappearance,
$\Delta P^{\rm T}_{\mu\mu}$ and $\Delta P^{\rm T}_{\tau \tau}$, are non-zero for asymmetric matter
profiles when $\delta\neq 0,\pi$, in contradiction with the frequent claim. This claim rests on a confusion between the proper and  improper T test, since the improper T test for disappearance vanishes, by construction. We also show that the proper T test for $\nu_e$, $\Delta P^{\rm T}_{ee}$,  vanishes for all matter profiles, because of the form of the Wolfenstein matter potential. Thus, we are in agreement with the claims made in Refs.~\cite{Kuo:1987km,Minakata:1999ze,Fishbane:2000kw, Yokomakura:2002av} for asymmetric matter profiles.\\

Proper CP changes neutrinos to anti-neutrinos and the matter profile to an anti-matter profile, which is achieved by changing $U \to U^*$. Similarly, we have shown that proper T reverses the propagation of neutrinos, which is also achieved by replacing $U \to U^*$. Therefore, CPT, which changes neutrinos to anti-neutrinos, matter to anti-matter and reverses the direction of propagation through the matter profile, is achieved by replacing $U \to U^*$ twice, which gives no change as expected. 
For any matter profile and any flavor pair:
$\Delta P^{\rm T}_{\alpha\beta}=\Delta P^{\rm CP}_{\alpha\beta}$
(both purely intrinsic, vanishing iff $\delta=0,\pi$), and
$\Delta P^{\rm CPT}_{\alpha\beta}=0$ for all $\delta$. The only flavor pair that satisfies the first of these trivially is $(ee)$: $\Delta P^{\rm T}_{ee}=0=\Delta P^{\rm CP}_{ee}$, for the other disappearance channels $\Delta P^{\rm T}_{\mu \mu}\neq 0$ and  $\Delta P^{\rm T}_{\tau \tau}\neq 0$. $\nu_e$ is special because of the form of the Wolfenstein matter potential.\\

To explore the physics of T-violation in $\nu_\mu$ disappearance, we consider a two layer experiment where the density of the two layers is different so that we have an asymmetric matter profile. We find that the maximum T-violation occurs when one of the layers has a density times energy close to the solar resonance
($\rho E_\nu \sim 0.4$ g/cm$^3$) and the other close to the atmospheric resonance ($\rho E_\nu \sim 30$ g/cm$^3$) with lengths comparable to the solar oscillation length ($\sim$15,000~km/GeV) and the atmospheric oscillation length ($\sim$500 km/GeV). In this setup the T-violation can be exceptionally large with 
\begin{align}
    \Delta P^{\rm T}_{\mu \mu} \sim 0.8 \times \sin \delta \,.
\end{align}
For normal ordering this effect occurs in the muon neutrino disappearance channel whereas for inverted ordering the effect is in the muon anti-neutrino disappearance channel.
Clearly such a setup is not possible with both  the neutrino source and detector on  or within a few kilometers of the Earth's surface.\\

We have also explored whether the matter profile asymmetry in DUNE and Hyper-Kamiokande is large enough so that the T-violation in $\nu_\mu$ disappearance is measurable. At an energy of 0.5 GeV for DUNE and 0.1 GeV for Hyper-Kamiokande the T-violating coefficient, $B$, is 0.01\% and gets smaller at higher energy. Thus, T-violation is unmeasurable in these experiments in the $\nu_\mu$ disappearance channel.\\

To obtain an observable effect in $\nu_\mu$ disappearance we need to consider
longer path lengths than available on the Earth's surface. We consider two such futuristic cases: (1) a neutrino source on the Earth pointing at a detector in geostationary orbit on the other side of the Earth and (2) a neutrino source on the Moon and several detectors on Earth. Both setups give sizable effects with
   $ \Delta P^{\rm T}_{\mu \mu} \sim (0.3 - 0.6) \times \sin \delta $.
   Given the commercialization of space, neither of these futuristic possibilities seem as far off as they once would have seemed. Exploring the Earth's interior with a $\nu_\mu$ beam using (1) may even have a financial incentive given its sensitivity to the Earth's structure.\\

In closing, proper T-violation in $\nu_\mu$ and $\nu_\tau$ disappearance in an asymmetric matter profile 
can be probed if the matter profiles are sufficiently asymmetric with a very long baseline\footnote{Our title is a violation of Hinchliffe's rule.}.

\section*{Acknowledgments}
\noindent
GB is supported by the Spanish grants  PID2020-113775GB-I00 
(AEI/10.13039/501100011033), and by the European ITN project HIDDeN (H2020-MSCA-ITN-2019/860881-HIDDeN).
This work was produced by Fermi Forward Discovery Group, LLC under Contract No. 89243024CSC000002 with the U.S. Department of Energy, Office of Science, Office of High Energy Physics. Publisher acknowledges the U.S. Government license to provide public access under the DOE Public Access Plan (www.energy.gov/doe-public-access-plan). ASV acknowledges support from the Spanish Grant FPU23/01408 funded by MICIU and thanks Fermilab for hospitality during a scientific visit. SJP thanks Peter Denton for discussions on this topic. No A.I. was used in this paper, only human intelligence was used.
\\.

\clearpage
\appendix

\section{Further discussion on the Reversal Identity}
\label{app:reverse}

In this appendix, we have generalized the discussion found in Ref.~\cite{Fishbane:2000kw} using multiple layers of uniform matter density, which we find more transparent than the discussion given in Ref.~\cite{Akhmedov:2001kd}. We start with a single layer of uniform density and show how proper CP, T and CPT act for this scenario and then we generalize for multiple layers with different densities. Before considering the single layer scenario, we want to use the re-phasing convention of the mixing matrix in vacuum, $U$, such that it is real when the Jarlskog invariant is zero and similarly for the mixing matrix, in matter, $V$. Consequently,  $U \to U^*$ also implies $V \to V^*$ and when there is no T and CP violation
both $U$ and $V$ are real matrices. This implies that in vacuum, $V=U$, with no re-phasing factors. \\

\underline{One Layer of Uniform Density:}\\[2mm]
Starting with the forward propagation of neutrinos in this uniform layer of length $L$ in matter, the evolution matrix is obtained from the Hamiltonian, $H$, giving
\begin{align}
S_\mathbf{1}[U] &= V \text{\bf ~diag}(1,e^{-i\Delta \lambda_{21}L/2E_\nu}, e^{-i\Delta \lambda_{31}L/2E_\nu})~  V^\dagger %=S^T_\mathbf{1}[U^*] 
\end{align}
that satisfies unitarity, $S_\mathbf{1}[U] S^\dagger_\mathbf{1}[U]=1$. For a constant density layer, $S_\mathbf{1}[U]=(S_\mathbf{1}[U^*])^T$.
For the probabilities of neutrinos propagating forward in matter, we have
\begin{align}
\label{eq:P_betaalpha_single_layer}
    P^\rightarrow_{\alpha \beta}[U] &=|S_\mathbf{1}[U]_{\beta \alpha}|^2 \notag \\
    P^\rightarrow_{\beta \alpha}[U] &=|S_\mathbf{1}[U]_{ \alpha \beta}|^2 =|(S^T_\mathbf{1}[U^*])_{\alpha \beta}|^2= P^\rightarrow_{\alpha \beta}[U^*]
    \end{align}
    which are not equal unless the  Jarlskog invariant is zero, $J=0$. \\

The Proper CP transformed  evolution matrix is obtained using the Hamiltonian $H^*$:
\begin{align}
S_\mathbf{CP}[U] &= V^* \text{\bf ~diag}(1,e^{-i\Delta \lambda_{21}L/2E_\nu}, e^{-i\Delta \lambda_{31}L/2E_\nu})~  V^T 
%= S^T_\mathbf{CP}[U^*] 
\\
&= S_\mathbf{1}[U^*] \,. \notag
\end{align}
 Again, unitarity is satisfied, $S_\mathbf{CP}[U] S^\dagger _\mathbf{CP}[U]=1$.
For the  probabilities of anti-neutrinos propagating in anti-matter, we have
\begin{align}
    \widehat{\overline{P}}\,^\rightarrow_{\alpha \beta}[U] &=|S_\mathbf{CP}[U]_{ \beta \alpha}|^2=  |S_\mathbf{1}[U^*]_{ \beta \alpha}|^2 =P^\rightarrow_{\alpha \beta}[U^*].
\end{align}
This is the CP Identity for a single layer.\\

The Proper T transformed  evolution matrix is obtained using the Hamiltonian,\\ $-H^T$:
\begin{align}
S_\mathbf{T}[U] &= V^* \text{\bf ~diag}(1,e^{+i\Delta \lambda_{21}L/2E_\nu}, e^{+i\Delta \lambda_{31}L/2E_\nu})~  V^T  
%= S^T_\mathbf{T}[U^*]
\\
&= S^*_\mathbf{1}[U]  = S^\dagger_\mathbf{1}[U^*]. \notag
\end{align}
 Again, unitarity is satisfied, $S_\mathbf{T}[U] S^\dagger _\mathbf{T}[U]=1$. However, $S_\mathbf{1}[U] S_\mathbf{T}[U]\neq 1$ unless $V$ and hence $U$ are real, i.e., if there is no intrinsic T-violation, $J=0$.
For the T-reversed probabilities of neutrinos propagating in matter, we have
\begin{align}
\label{eq:reversal id single layer}
    P^\leftarrow_{\alpha \beta}[U] &=|S_\mathbf{T}[U]_{ \alpha \beta }|^2=  |S_\mathbf{1}[U^*]_{ \beta \alpha}|^2 =P^\rightarrow_{\alpha \beta}[U^*].
\end{align}
This is the {\bf reversal identity} for a single layer. Since this uniform layer is a symmetric matter profile,  $P^\leftarrow_{\alpha \beta}[U]=P^\rightarrow_{\beta \alpha}[U]$ as required, since both eqs.~\eqref{eq:P_betaalpha_single_layer} and \eqref{eq:reversal id single layer} are equal to $P^\rightarrow_{\alpha \beta}[U^*]$.
\\

The Proper CPT transformed  evolution matrix is obtained from the Hamiltonian $-H^\dagger = -H$:
\begin{align}
S_\mathbf{CPT}[U] &= V \text{\bf ~diag}(1,e^{+i\Delta \lambda_{21}L/2E_\nu}, e^{+i\Delta \lambda_{31}L/2E_\nu})~  V^\dagger 
%= S^T_\mathbf{CPT}[U^*] 
\\
&=S^\dagger_\mathbf{1}[U] \, .\notag
\end{align}
Again $S_\mathbf{CPT}[U] S^\dagger _\mathbf{CPT}[U]=1$, and $S_\mathbf{1}[U] S_\mathbf{CPT}[U] = 1$, for all values of the Jarlskog invariant,
as expected, since CPT is conserved.
For the  reversed probabilities of anti-neutrinos in anti-matter, we have
\begin{align}
    \widehat{\overline{P}}\,^\leftarrow_{\alpha \beta}[U] &=|S_\mathbf{CPT}[U]_{ \alpha \beta}|^2=  |S^*_\mathbf{1}[U]_{ \beta \alpha}|^2 =P^\rightarrow_{\alpha \beta}[U]\, ,
\end{align}
as required by CPT conservation. At the probability level, since both T-reversal and CP require $U \to U^*$, CPT is clearly satisfied.\\

\underline{Multi-Layer Density Profile:}\\
For a multi-layer density profile, with different densities for each layer, to obtain the evolution matrices for $\{ 1, \mathrm{CP}, \mathrm{T}, \mathrm{CPT} \}$, we just need to multiply the appropriate single layer evolution matrices
in the right order. Note that T and CPT are reversed with respect to both $\mathbf{1}$ and CP.
\begin{align}
    S_\mathbf{1}[U](1, \ldots,m)=&S_\mathbf{1}[U,W_m] \ldots S_\mathbf{1}[U,W_1] \notag \\
    S_\mathbf{CP}[U](1, \ldots,m)=&S_\mathbf{CP}[U,W_m] \ldots S_\mathbf{CP}[U,W_1]  \\[3mm]
    S_\mathbf{T}[U](1, \ldots,m)=&S_\mathbf{T}[U,W_1] \ldots S_\mathbf{T}[U,W_m] \notag \\
    S_\mathbf{CPT}[U](1, \ldots,m)=&S_\mathbf{CPT}[U,W_1] \ldots S_\mathbf{CPT}[U,W_m]
    \end{align}
    \begin{enumerate}
\item 
For CP: using that $S_\mathbf{CP}[U,W_k]=S_\mathbf{1}[U^*,W_k]$ then $S_\mathbf{CP}[U](1, \ldots,m)= S_\mathbf{1}[U^*](1, \ldots,m)$ and hence
$$ \widehat{\overline{P}}\,^\rightarrow_{\alpha \beta}[U,W] =  P^\rightarrow_{\alpha \beta}[U^*,W]\,, $$
the CP Identity.
\item 
For T: using that $S_\mathbf{T}[U,W_k]=S^\dagger_\mathbf{1}[U^*,W_k]$ then $S_\mathbf{T}[U](1, \ldots,m)= S^\dagger_\mathbf{1}[U^*](1, \ldots,m)$ 
and $S_\mathbf{1}[U](1, \ldots,m)S_\mathbf{T}[U](1, \ldots,m) \neq 1$, as was also the case for a single layer. For the probabilities we obtain
$$ P\,^\leftarrow_{\alpha \beta}[U,W] =  P^\rightarrow_{\alpha \beta}[U^*,W] \,, $$ the Reversal Identity.
\item 
For CPT: using that $S_\mathbf{CPT}[U,W_k]=S^\dagger_\mathbf{1}[U,W_k]$ then $S_\mathbf{CPT}[U](1, \ldots,m)= S^\dagger_\mathbf{1}[U](1, \ldots,m)$ 
and $S_\mathbf{1}[U](1, \ldots,m)S_\mathbf{CPT}[U](1, \ldots,m) =1 $, as was also the case for a single layer. For the probabilities we obtain
$$ \widehat{\overline{P}}\,^\leftarrow_{\alpha \beta}[U,W]=  P^\rightarrow_{\alpha \beta}[U,W] \,, $$ the CPT Identity.
    \end{enumerate}

In this paper, we are primarily interested in the probabilities so any distinction between Wigner reversal, implemented by the change $W \to W_{\rm rev}$, and T-reversal is beyond the scope of this paper as at the probability level there is no distinction.
\newpage

\section{Oscillation probability coefficients}
\label{app:coefficients}
The oscillation probabilities are given by
\begin{align}
% \label{eq:P_alpha beta}[
P\,^\rightarrow_{\alpha\beta}[U,W] &=|S_{\beta \alpha}|^2= P\,^\leftarrow_{\alpha\beta}[U^*,W] = P\,^\rightarrow_{\beta\alpha}[U^*,W_{rev}] \notag \\
&= A_{\alpha\beta}
+ B_{\alpha\beta}\sin\delta
+ C_{\alpha\beta}\cos\delta
+ D_{\alpha\beta}\sin 2\delta
+ E_{\alpha\beta}\cos 2\delta,
\end{align}
where eq. \eqref{eq:Smatrix} can be rewritten as
\begin{align}
    S= &\left(\begin{array}{ccc}
    \widetilde{S}_{11}  
    & \widetilde{S}_{12 }c_{23}
    & -\widetilde{S}_{12} s_{23}\\[3mm] 
    \widetilde{S}_{21} c_{23}
    &\widetilde{S}_{22} c^2_{23} + \widetilde{S}_{33}s^2_{23}
    & (\widetilde{S}_{33}-\widetilde{S}_{22} )c_{23}s_{23}  \\[3mm]
    -\widetilde{S}_{21} s_{23}
    &    (\widetilde{S}_{33}-\widetilde{S}_{22} )c_{23}s_{23} 
    &\widetilde{S}_{33}c^2_{23}+\widetilde{S}_{22} s^2_{23} 
    \end{array} \right) \notag \\[2mm]
%\end{align}
%\begin{align}
     +\cos \delta  &\left(\begin{array}{ccc}
    0 
    & \widetilde{S}_{13 } s_{23}  
    & \widetilde{S}_{ 13 }c_{23}\\[3mm] 
    \widetilde{S}_{31}  s_{23} 
    &(\widetilde{S}_{23}+ \widetilde{S}_{32}) c_{23}s_{23}
    & \widetilde{S}_{23}c^2_{23}  - \widetilde{S}_{32} s^2_{23} 
    \\[3mm]
    \widetilde{S}_{31} c_{23}
    &  \widetilde{S}_{32}c^2_{23} - \widetilde{S}_{23}  s^2_{23}  
    &-(\widetilde{S}_{23} + \widetilde{S}_{32})c_{23}s_{23}
    \end{array} \right) \notag \\[2mm]
%\end{align}
%\begin{align}
    + i\sin \delta  &\left(\begin{array}{ccc}
    0 
    & -\widetilde{S}_{13 } s_{23}  
    & -\widetilde{S}_{ 13 }c_{23}\\[3mm] 
    \widetilde{S}_{31}  s_{23} 
    &( \widetilde{S}_{32} - \widetilde{S}_{23})c_{23}s_{23}
    & -(\widetilde{S}_{23}c^2_{23}  + \widetilde{S}_{32} s^2_{23} )
    \\[3mm]
    \widetilde{S}_{31} c_{23}
    &  \widetilde{S}_{32}c^2_{23} + \widetilde{S}_{23}  s^2_{23}  
    &-(\widetilde{S}_{32} - \widetilde{S}_{23})c_{23}s_{23}
    \end{array} \right) \,.
\end{align}
%}
$S_{\beta \alpha}$ for $\alpha \neq \beta$ is obtained by interchanging the subscripts on $\widetilde{S}$ and flipping the sign of $\delta$. $S_{\tau \tau}$ can be obtained from $S_{\mu \mu}$ by interchanging $c^2_{23} \leftrightarrow s^2_{23}$ and flipping the sign of $s_{23}c_{23}$. For each channel, evaluating $P\,^\rightarrow_{\alpha\beta}$ at $\delta =(0,~\pm\pi/4,~\pm \pi/2,~\pi)$ gives more than enough information to determine $A,\ldots,E$ for that channel.\\

Using the following identity
\begin{align}
    |~X+Y\cos \delta +i Z\sin \delta~|^2 &= |X|^2+\frac1{2}(|Y|^2+|Z|^2) \notag \\
    ~+~ 2 \Im [XZ^*]\sin \delta ~+~ 2 \Re[XY^*]\cos \delta & ~+~ \Im [YZ^*]\sin 2\delta ~+~\frac1{2}(|Y|^2-|Z|^2)\cos 2\delta \notag
\end{align}
where $X$, $Y$ and $Z$ are complex quantities. \\

The $\nu_e$ probabilities are given in Table \ref{tab:Pe} whereas the  probabilities only involving $\nu_\mu$ and $\nu_\tau$ are given in Table~\ref{tab:Pmt}. The results in these two tables are equal to those given in Appendix A of Ref.~\cite{Yokomakura:2002av}.

\begin{table}[b]
    \begin{tabular}{c|c|c|c|c|}
    & $\mathbf{P\,^\rightarrow_{e \mu}=|S_{\mu e}|^2}$ &  $\mathbf{P\,^\rightarrow_{e \tau}=|S_{\tau e}|^2}$ & $\mathbf{P\,^\rightarrow_{\mu e}=|S_{e\mu}|^2}$ & $\mathbf{P\,^\rightarrow_{\tau e}=|S_{e\tau}|^2}$ \\[2mm]
    \hline &&&&\\[-2mm]
    $A$ 
    &  $|\widetilde{S}_{21}|^2 c_{23}^2 + |\widetilde{S}_{31}|^2 s_{23}^2$
     &  $|\widetilde{S}_{21}|^2 s_{23}^2 + |\widetilde{S}_{31}|^2 c_{23}^2$
     & $|\widetilde{S}_{12}|^2 c_{23}^2 + |\widetilde{S}_{13}|^2 s_{23}^2$
     & $|\widetilde{S}_{12}|^2 s_{23}^2 + |\widetilde{S}_{13}|^2 c_{23}^2$\\[2mm]
     $B$
     & $2 \Im \left[ \widetilde{S}_{21} \widetilde{S}^*_{31} \right] c_{23} s_{23}$
     & $-B_{e \mu}$
     & $-2 \Im \left[ \widetilde{S}_{12} \widetilde{S}^*_{13} \right] c_{23} s_{23}$
     & $-B_{\mu e}$ \\[2mm]
     $C$
     & $ 2 \Re \left[ \widetilde{S}_{21} \widetilde{S}^*_{31} \right] c_{23} s_{23}$
     & $-C_{e\mu}$
     & $\phantom{-} 2 \Re \left[ \widetilde{S}_{12} \widetilde{S}^*_{13} \right] c_{23} s_{23}$
     & $-C_{\mu e}$
    \\ \hline  
    \end{tabular}
    \caption{Together with $\mathbf{P\,^\rightarrow_{ee}=|\widetilde{S}_{11}|^2}$ give all the probabilities involving $\nu_e$. The coefficients $D=E=0$ for these channels.
    It is easy to see that unitarity,  $P\,^\rightarrow_{ee}+P\,^\rightarrow_{e\mu}+P\,^\rightarrow_{e\tau}=
    P\,^\rightarrow_{ee}+P\,^\rightarrow_{\mu e}+P\,^\rightarrow_{\tau e}=1$, is satisfied as
    $\widetilde{S}$ is unitary. For $X=(A,C)$ note that $X_{e\mu} \neq X_{\mu e}$, $X_{e\tau} \neq X_{\tau e}$ and  $B_{e\mu} \neq -B_{\mu e}$, $B_{e\tau} \neq -B_{\tau e}$ unless $\widetilde{S}$ is symmetric which requires the matter profile to be symmetric. 
    }
    \label{tab:Pe}
\end{table}

\begin{table}[b]
\hspace*{-1cm}
    \begin{tabular}{c|c|c|}
    & $\mathbf{P\,^\rightarrow_{\mu \mu}=|S_{\mu \mu}|^2}$ &  $\mathbf{P\,^\rightarrow_{\tau \mu}=|S_{\mu \tau}|^2}$ \\[2mm]
    \hline &&\\[-2mm]
    $A$ 
    & $ \bigl|\widetilde{S}_{22} c^2_{23}+ \widetilde{S}_{33}s^2_{23}\bigr|^2
             + \bigl(|\widetilde{S}_{23}|^2 + |\widetilde{S}_{32}|^2\bigr)s^2_{23}c^2_{23}$
    & $|\widetilde{S}_{33} - \widetilde{S}_{22}|^2 c_{23}^2 s_{23}^2 +|\widetilde{S}_{32}|^2 s_{23}^4 + |\widetilde{S}_{23}|^2 c_{23}^4$ 
             \\[2mm]
    $B$ 
    & $2\,\mathrm{Im}
\!\left[\bigl(\widetilde{S}_{22} c^2_{23}+ \widetilde{S}_{33}s^2_{23}\bigr)
\bigl(\widetilde{S}_{32}-\widetilde{S}_{23}\bigr)^*\right]c_{23}s_{23}$
    & $- 2 \Im \left[ (\widetilde{S}_{33} - \widetilde{S}_{22} )
    (\widetilde{S}_{32} s^2_{23} + \widetilde{S}_{23} c^2_{23})^*  \right] c_{23} s_{23}$
    \\[2mm]
    $C$ 
    & $2\,\mathrm{Re}
             \!\left[\bigl(\widetilde{S}_{22} c^2_{23}+ \widetilde{S}_{33}s^2_{23}\bigr)
             \bigl(\widetilde{S}_{32}+\widetilde{S}_{23}\bigr)^*\right]c_{23}s_{23}$
    & $ - 2 \Re \left[ (\widetilde{S}_{33} - \widetilde{S}_{22} )
    (\widetilde{S}_{32} s^2_{23} - \widetilde{S}_{23} c^2_{23})^*  \right] c_{23} s_{23} $
    \\[2mm]
    $D$
    & $2\,\mathrm{Im}\bigl[\widetilde{S}_{23}\widetilde{S}^*_{32}\bigr]c^2_{23}s^2_{23}$
    & $-D_{\mu \mu}$
    \\[2mm]
    $E$
    & $2\,\mathrm{Re}\bigl[\widetilde{S}_{23}\widetilde{S}^*_{32}\bigr]c^2_{23}s^2_{23}$
    & $-E_{\mu \mu}$\\[2mm]
    \hline \hline \hline && \\[-2mm]
    & $\mathbf{P\,^\rightarrow_{\mu \tau}=|S_{\tau \mu}|^2}$ &  $\mathbf{P\,^\rightarrow_{\tau \tau}=|S_{\tau \tau}|^2}$ \\[2mm]
    \hline &&\\[-2mm]
    $A$ 
    & $|\widetilde{S}_{33} - \widetilde{S}_{22}|^2 c_{23}^2 s_{23}^2 +|\widetilde{S}_{32}|^2 c_{23}^4 + |\widetilde{S}_{23}|^2 s_{23}^4 $
    & $\bigl|\widetilde{S}_{33} c^2_{23}+ \widetilde{S}_{22}s^2_{23}\bigr|^2
             + \bigl(|\widetilde{S}_{23}|^2 + |\widetilde{S}_{32}|^2\bigr)s^2_{23}c^2_{23}$
             \\[2mm]
    $B$ 
    & $+2 \Im \left[ (\widetilde{S}_{33} - \widetilde{S}_{22} )(\widetilde{S}_{32} c^2_{23} + \widetilde{S}_{23} s^2_{23})
    ^* 
    \right] c_{23} s_{23}$
    & $-2 \Im \left[ 
     (\widetilde{S}_{33} c^2_{23}+ \widetilde{S}_{22}s^2_{23})
     (\widetilde{S}_{32} - \widetilde{S}_{23})^*  \right] c_{23} s_{23} $
    \\[2mm]
    $C$ 
    & $+2 \Re \left[ (\widetilde{S}_{33} - \widetilde{S}_{22} ) 
    (\widetilde{S}_{32} c^2_{23} - \widetilde{S}_{23} s^2_{23})^* \right] c_{23} s_{23}$
    & $-2 \Re \left[ (\widetilde{S}_{33} c^2_{23}+ \widetilde{S}_{22}s^2_{23}) (\widetilde{S}_{32} + \widetilde{S}_{23})^* \right] c_{23} s_{23}$
    \\[2mm]
    $D$
    & $-D_{\mu \mu}$
    & $D_{\mu \mu}$
    \\[2mm]
    $E$
    & $-E_{\mu \mu}$
    & $E_{\mu \mu}$\\[2mm]
     \hline
    \end{tabular}
    \caption{The disappearance and appearance probabilites only involving $\nu_\mu$ and $\nu_\tau$. It is easy to check that unitarity,  $P\,^\rightarrow_{\mu \mu}+P\,^\rightarrow_{\mu \tau}+P\,^\rightarrow_{\mu e}=
    P\,^\rightarrow_{\tau \tau}+P\,^\rightarrow_{\tau \mu }+P\,^\rightarrow_{\tau e}=1$, is satisfied as $\widetilde{S}$ is unitary. The T-violating coefficients, $B$ and $D$, in $P\,^\rightarrow_{\mu \mu}$ and $P\,^\rightarrow_{\tau \tau}$ are non-zero, unless
    $\widetilde{S}$ is symmetric which requires the matter profile to be symmetric. 
    }
    \label{tab:Pmt}
\end{table}

\clearpage

Due to unitarity, the rows and columns of $A$ add to 1, whereas the row and columns of the rest of the coefficients add to 0:
\begin{align}
A_{ee} &= 1- A_{e\mu} -A_{e \tau}= 1- A_{\mu e} -A_{ \tau e}, \notag \\
A_{\mu \mu} &= 1- A_{\mu e} -A_{\mu \tau}= 1- A_{e\mu} -A_{ \tau \mu}, \notag \\
A_{\tau \tau } &= 1- A_{\tau \mu} -A_{ \tau e}= 1- A_{\mu \tau} -A_{e \tau},\\[2mm]
B_{ee}&=-B_{e \mu}- B_{e \tau} = -B_{ \mu e } - B_{ \tau e }=0, \notag \\
B_{\mu \mu} &=- B_{e \mu}-B_{ \tau \mu} = -B_{ \mu e} -B_{\mu \tau}, \notag  \\
B_{\tau \tau} &=-B_{\tau e}-B_{ \tau \mu} = -B_{ e \tau} - B_{\mu \tau},  \\[2mm]
C_{ee}&=-C_{e \mu} - C_{e \tau} =-C_{ \mu e}- C_{ \tau e } =0, \notag \\
C_{\mu \mu} &=-C_{e \mu}-C_{ \tau \mu} = -C_{ \mu e} -C_{\mu \tau}, \notag \\
C_{\tau \tau} &=-C_{\tau e}-C_{ \tau \mu} = -C_{ e \tau} -C_{\mu \tau}, \\[2mm]
D_{\mu \mu} &= -D_{\mu \tau}= -D_{\tau \mu} = D_{\tau \tau},\\
E_{\mu \mu} &= -E_{\mu \tau}= -E_{\tau \mu} = E_{\tau \tau}.
\end{align}

Note that for antineutrinos, we write the oscillation probability as
\begin{align}
\overline{P}\,^\rightarrow_{\alpha\beta}[U,W] &= P\,^\rightarrow_{\alpha\beta}[U^*,-W] \notag \\
&=\overline A_{\alpha \beta}   - \overline B_{\alpha \beta} \sin \delta  + \overline C_{\alpha \beta} \cos \delta - \overline D_{\alpha \beta} \sin 2\delta + \overline E_{\alpha \beta} \cos 2\delta .
\end{align}
Note the minus sign on the T/CP-odd coefficients. With this choice, we can write $\overline X_{\alpha \beta} = X_{\alpha \beta} ( -W)$, which means that we need to re-solve the Schrodinger equation after flipping the sign of the Wolfenstein potential.

\clearpage

\section{CP, T and CPT cheat sheet}
\label{sec:cheat}
\definecolor{cpcol}{RGB}{220,50,47}
\definecolor{tcol}{RGB}{38,139,210}
\definecolor{cptcol}{RGB}{133,153,0}

\begin{table*}[!h]
\centering
\small
\begin{tabular}{cccc}
\toprule
Test & Vacuum & Symmetric matter \quad & Asymmetric matter \\
\midrule
$\Delta P^{\rm T}_{\alpha\beta}$ (prop.) 
& $2 B \sin\delta$
& $2 B \sin\delta$
& $2(B\sin\delta + D\sin2\delta)$ \\[1mm]

$\Delta P^{\rm CP}_{\alpha\beta}$ (prop.) 
& $2 B \sin\delta$
& $2 B \sin\delta$
& $2(B\sin\delta + D\sin2\delta)$ \\[1mm]

$\Delta P^{\rm CPT}_{\alpha\beta}$ (prop.)
& $0$ & $0$ & $0$ \\[2mm]
\toprule

$\Delta P^{\tau}_{\alpha\beta}$ (impr.) 
& $2 B \sin\delta$
& $2 B \sin\delta$
& {\footnotesize $(A_{\alpha\beta}-A_{\beta\alpha}) + (B_{\alpha\beta}-B_{\beta\alpha})\sin\delta $} \\ 
& 
& 
& {\footnotesize $+ (C_{\alpha\beta}-C_{\beta\alpha})\cos\delta +   (D_{\alpha\beta}-D_{\beta\alpha})\sin2\delta$} \\
& 
& 
& {\footnotesize $ + (E_{\alpha\beta}-E_{\beta\alpha})\cos2\delta $} \\[3mm]

$\Delta P^{\rm cp}_{\alpha\beta}$ (impr.) 
& $2 B \sin\delta$
& {\footnotesize $ (A-\overline A) + (B+\overline B)\sin\delta  $}
& {\footnotesize $(A-\overline A) + (B+\overline B)\sin\delta  $} \\ 
& 
& {\footnotesize $ +(C-\overline C)\cos\delta $}
& {\footnotesize $ +(C-\overline C)\cos\delta + (D+\overline D)\sin2\delta$} \\
& 
& {\footnotesize $ +  (E-\overline E)\cos2\delta$}
& {\footnotesize $  +  (E-\overline E)\cos2\delta$} \\[3mm]

$\Delta P^{\rm cp\tau}_{\alpha\beta}$ (impr.)
& $0$
& {\footnotesize $(A_{\alpha\beta}-\overline A_{\beta\alpha}) $}
& {\footnotesize $(A_{\alpha\beta}-\overline A_{\beta\alpha}) + (B_{\alpha\beta}+\overline B_{\beta\alpha})\sin\delta $} \\
& 
& {\footnotesize $+ (B_{\alpha\beta}+\overline B_{\beta\alpha})\sin\delta $}
& {\footnotesize $+ (C_{\alpha\beta}-\overline C_{\beta\alpha})\cos\delta $} \\
& 
& {\footnotesize $+ (C_{\alpha\beta}-\overline C_{\beta\alpha})\cos\delta    $}
& {\footnotesize $ + (D_{\alpha\beta}+\overline D_{\beta\alpha})\sin 2\delta  $} \\
& 
& {\footnotesize $ + (E_{\alpha\beta}-\overline E_{\beta\alpha})\cos2\delta$}
& {\footnotesize $ +  (E_{\alpha\beta}-\overline E_{\beta\alpha})\cos 2\delta $} \\[3mm]
\toprule

\end{tabular}
\caption{Proper and improper tests for all channels, $\nu_\alpha\to\nu_\beta$, for any $\alpha$ and $\beta$. Wherever the coefficients $X=A, B, C, D, E$ do not present subscripts, it is understood $X_{\alpha \beta}$.
 In vacuum, the proper and improper tests coincide for both CP and T,
  since $W=0$ ($\mathrm{cp}=\mathrm{CP}$) and the profile is trivially symmetric ($\tau=\mathrm{T}$).
  In symmetric matter, the presence of a material medium makes $\mathrm{cp}\neq\mathrm{CP}$, but $\tau=\mathrm{T}$ still holds, since reversing a
  symmetric profile leaves it unchanged.
  The improper $\tau$ test, $\Delta P^\tau_{\alpha\beta}$, acquires an
  extrinsic contribution in asymmetric matter and can be non-zero
  even at $\delta=0$, unlike the proper test $\Delta P^{\rm T}_{\alpha\beta}$
  which remains a clean probe of $\delta$ in all three regimes. Note that for symmetric matter (and vacuum), $B_{\alpha \alpha} = 0$ (only for disappearance channels) and $D_{\alpha \beta} = 0$ (for any channel).
$D$ and $E$ are zero for any channel with $\alpha$ and/or $\beta$ equal to $e$.}
\label{tab:appearance}
\end{table*}

\clearpage  

\section{Measuring T/CP-conserving terms}
\label{app:Cmax}
\begin{figure}[h]
    \centering
    \includegraphics[width=0.8\linewidth]{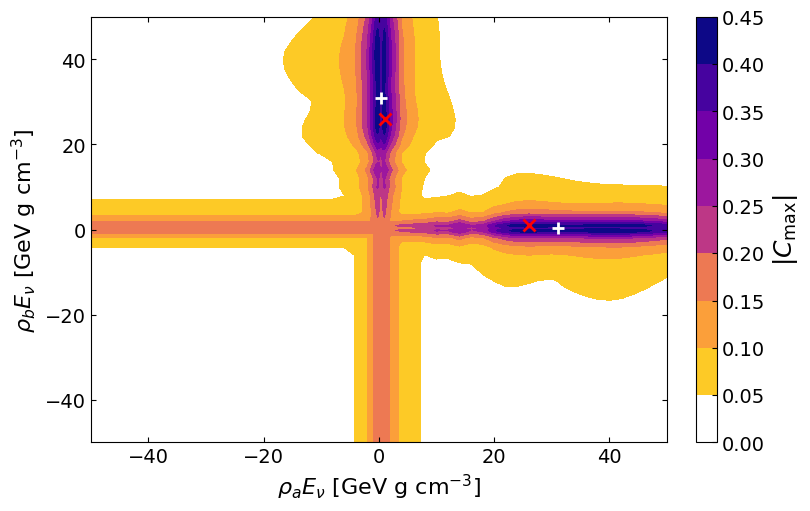}
    \caption{Absolute modulus of $C$, the term proportional to $\cos \delta$ in $P_{\mu \mu}$, as a function of $\rho E_{\nu}$ for each layer. Unlike $B$, this term does not vanish when $\rho_a = \rho_b$ or at low $\rho E_{\nu}$. The white plus signs indicate where the solar and atmospheric resonances are located, and the red crosses where the global maximum is found.}
    \label{fig:Cmax}
\end{figure}

Another feasible option to determine $\delta$ through measuring $P_{\mu \mu}$ is to enhance the T/CP-conserving terms $\propto \cos \delta$, that is, $C$. This possibility was already pointed out in Refs.~\cite{Kimura:2006hy, Kimura:2007mu} and it has been explored for DUNE and HK in Ref.~\cite{Denton:2023qmd}. Here we study it for an arbitrary two-layer system. In Fig.~\ref{fig:Cmax} we show that even for low $\rho E_\nu$ this effect can be quite large. \\

\begin{figure}
    \centering
    \includegraphics[width=\linewidth]{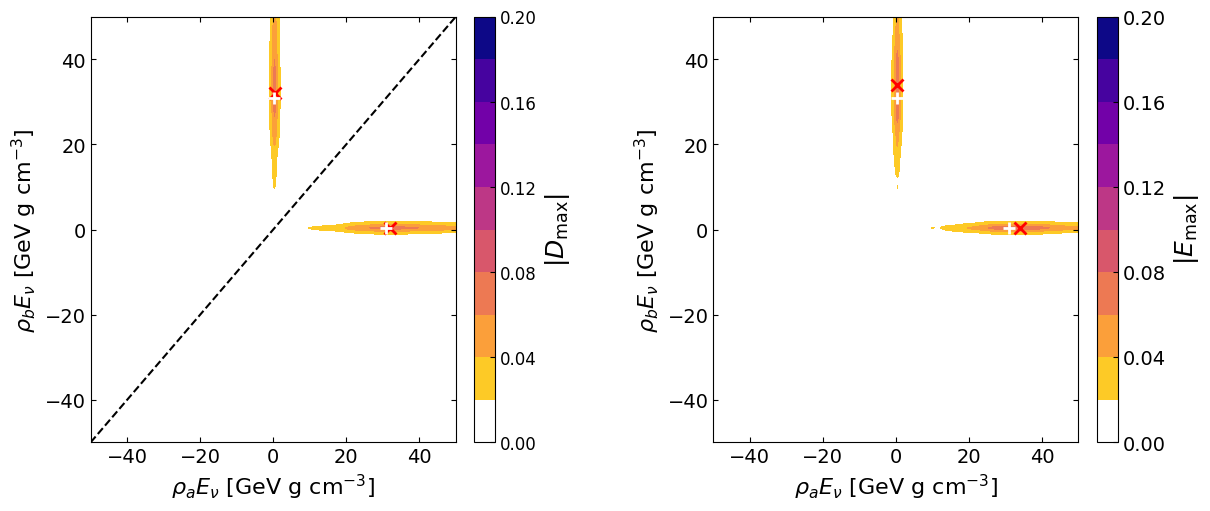}
    \caption{Absolute modulus of $D$, the term proportional to $\sin 2\delta$ in $P_{\mu \mu}$ (left panel) and $E$, the term proportional to $\cos 2\delta$ in $P_{\mu \mu}$ (right panel), both as a function of $\rho E_{\nu}$ for each layer. Note the change in scale with respect $B$ and $C$, these terms are generally subdominant in $P_{\mu \mu}$. The white plus signs indicate where the solar and atmospheric resonances are located, and the red crosses where the global maximum  is found.}
    \label{fig:DmaxEmax}
\end{figure}

For completeness, we show the size of the coefficients $D$ and $E$ that contribute to $P_{\mu \mu}$ in Fig.~\ref{fig:DmaxEmax}. $A$ is not shown since its maximum value is always $1$. Note the different scale in comparison with Figs.~\ref{fig:Bmax} and \ref{fig:Cmax}. This does not necessarily mean that both coefficients can be neglected when compared with $B$ and $C$, since all terms oscillate with the propagation lengths and we are just plotting their maximum value for \underline{any} pair $\{ L_a, L_b \}$. The pair of $\{ L_a, L_b \}$ that maximizes one of the coefficients must not be the same for all of them. In general, the interplay between all coefficients must be considered. However, to a first approximation, one could restrict to the coefficients $A, B$ and $C$.

\subsection{Symmetry under $a \leftrightarrow b$}
\label{app:a to b}
In Figs.~\ref{fig:Bmax}, \ref{fig:Cmax} and~\ref{fig:DmaxEmax}, the exchange $a \leftrightarrow b$ leaves invariant the absolute modulus of the given term of the probability, $|X|$. If we take a term of the probability for a given setup, $X_{\alpha \beta} (W)$, and we interchange $a \leftrightarrow b$, we end up with $X_{\alpha \beta} (W_{\rm rev})$\footnote{Note that the interchange of layers $a \leftrightarrow b$ is equivalent to apply proper $T$ and improper $\tau$ consecutively.}. Using eqs.~\eqref{eq:PWrev} and \eqref{eq:master}, $P\,^\rightarrow_{\alpha\beta}[U,W_{\rm rev}] =P\,^\leftarrow_{\beta\alpha}[U,W]
= P\,^\rightarrow_{\beta\alpha}[U^*,W]
$, we find that  
\begin{equation}
    X_{\alpha \beta} (W_{\rm rev}) = \pm X_{\beta \alpha} (W),
\end{equation}
where the $+1$ for $X = A, C, E$ (T/CP-conserving terms, even under $U \to U^*$) and $-1$ for $X =B, D$ (T/CP-violating terms, odd under $U \to U^*$), therefore for disappearance $|X_{\alpha \alpha}(W_{\rm rev})| = | X_{\alpha \alpha} (W) | $.  For a general matter profile $X_{\beta \alpha}(W)$ is not directly related to $X_{\alpha \beta}(W)$ (see Appendix~\ref{app:coefficients} for all relations). \\
\clearpage

\bibliography{biblio}

@article{Akhmedov:2001kd,
    author = "Akhmedov, Evgeny K. and Huber, Patrick and Lindner, Manfred and Ohlsson, Tommy",
    title = "{T violation in neutrino oscillations in matter}",
    eprint = "hep-ph/0105029",
    archivePrefix = "arXiv",
    reportNumber = "TUM-HEP-413-01, MPI-PHT-01-11",
    doi = "10.1016/S0550-3213(01)00261-9",
    journal = "Nucl. Phys. B",
    volume = "608",
    pages = "394--422",
    year = "2001"
}

@article{Akhmedov:2004ve,
    author = "Akhmedov, Evgeny K.",
    editor = {Bergstr{\"o}m, L. and Botner, O. and Carlson, P. and Hulth, P. O. and Ohlsson, T.},
    title = "{Three-flavor effects and CP- and T-violation in neutrino oscillations}",
    eprint = "hep-ph/0412029",
    archivePrefix = "arXiv",
    doi = "10.1088/0031-8949/2005/T121/009",
    journal = "Phys. Scripta T",
    volume = "121",
    pages = "65--71",
    year = "2005"
}

@article{Bitter:2024eka,
    author = "Bitter, Olivia M. and de Gouv{\^e}a, Andr{\'e} and Kelly, Kevin J.",
    title = "{T-invariance violation in neutrino oscillations and matter effects}",
    eprint = "2412.13287",
    archivePrefix = "arXiv",
    primaryClass = "hep-ph",
    reportNumber = "MI-HET-846",
    doi = "10.1103/PhysRevD.111.055023",
    journal = "Phys. Rev. D",
    volume = "111",
    number = "5",
    pages = "055023",
    year = "2025"
}

@article{Denton:2023qmd,
    author = "Denton, Peter B.",
    title = "{Probing CP Violation with Neutrino Disappearance Alone}",
    eprint = "2309.03262",
    archivePrefix = "arXiv",
    primaryClass = "hep-ph",
    doi = "10.1103/PhysRevLett.133.031801",
    journal = "Phys. Rev. Lett.",
    volume = "133",
    number = "3",
    pages = "031801",
    year = "2024"
}

@article{Schwetz:2021thj,
    author = "Schwetz, Thomas and Segarra, Alejandro",
    title = "{T violation in nonstandard neutrino oscillation scenarios}",
    eprint = "2112.08801",
    archivePrefix = "arXiv",
    primaryClass = "hep-ph",
    doi = "10.1103/PhysRevD.105.055001",
    journal = "Phys. Rev. D",
    volume = "105",
    number = "5",
    pages = "055001",
    year = "2022"
}

@article{Blennow:2013rca,
    author = "Blennow, Mattias and Smirnov, Alexei Yu.",
    title = "{Neutrino propagation in matter}",
    eprint = "1306.2903",
    archivePrefix = "arXiv",
    primaryClass = "hep-ph",
    doi = "10.1155/2013/972485",
    journal = "Adv. High Energy Phys.",
    volume = "2013",
    pages = "972485",
    year = "2013"
}

@article{Fishbane:2000kw,
    author = "Fishbane, Paul M. and Kaus, Peter",
    title = "{Neutrino propagation in matter and CP violation}",
    eprint = "hep-ph/0012088",
    archivePrefix = "arXiv",
    doi = "10.1016/S0370-2693(01)00287-8",
    journal = "Phys. Lett. B",
    volume = "506",
    pages = "275--282",
    year = "2001"
}

@article{Yokomakura:2002av,
    author = "Yokomakura, Hidekazu and Kimura, Keiichi and Takamura, Akira",
    title = "{Overall feature of CP dependence for neutrino oscillation probability in arbitrary matter profile}",
    eprint = "hep-ph/0207174",
    archivePrefix = "arXiv",
    reportNumber = "DPNU-02-21",
    doi = "10.1016/S0370-2693(02)02545-5",
    journal = "Phys. Lett. B",
    volume = "544",
    pages = "286--294",
    year = "2002"
}

@article{Minakata:1999ze,
    author = "Minakata, Hisakazu and Watanabe, Shinji",
    title = "{Solar neutrinos and leptonic CP violation}",
    eprint = "hep-ph/9906530",
    archivePrefix = "arXiv",
    reportNumber = "TMUP-HEL-9908, WU-HEP-99-4",
    doi = "10.1016/S0370-2693(99)01224-1",
    journal = "Phys. Lett. B",
    volume = "468",
    pages = "256--260",
    year = "1999"
}

@article{Kuo:1987km,
    author = "Kuo, Tzee-Ke and Pantaleone, James T.",
    title = "{$T$ Nonconservation in Three Neutrino Oscillations}",
    reportNumber = "PURD-TH-87-12",
    doi = "10.1016/0370-2693(87)90688-5",
    journal = "Phys. Lett. B",
    volume = "198",
    pages = "406",
    year = "1987"
}

@article{Wolfenstein:1977ue,
    author = "Wolfenstein, L.",
    title = "{Neutrino Oscillations in Matter}",
    reportNumber = "COO-3066-102",
    doi = "10.1103/PhysRevD.17.2369",
    journal = "Phys. Rev. D",
    volume = "17",
    pages = "2369--2374",
    year = "1978"
}

@article{Roe:2017zdw,
    author = "Roe, Byron",
    title = "{Matter density versus distance for the neutrino beam from Fermilab to Lead, South Dakota, and comparison of oscillations with variable and constant density}",
    eprint = "1707.02322",
    archivePrefix = "arXiv",
    primaryClass = "hep-ex",
    reportNumber = "DUNE-DOC-3681-V2",
    doi = "10.1103/PhysRevD.95.113004",
    journal = "Phys. Rev. D",
    volume = "95",
    number = "11",
    pages = "113004",
    year = "2017"
}

@article{Kelly:2018kmb,
    author = "Kelly, Kevin J. and Parke, Stephen J.",
    title = "{Matter Density Profile Shape Effects at DUNE}",
    eprint = "1802.06784",
    archivePrefix = "arXiv",
    primaryClass = "hep-ph",
    reportNumber = "NUHEP-TH-18-02, NUHEP-TH/18-02, FERMILAB-PUB-18-019-T",
    doi = "10.1103/PhysRevD.98.015025",
    journal = "Phys. Rev. D",
    volume = "98",
    number = "1",
    pages = "015025",
    year = "2018"
}

@article{Hagiwara:2011kw,
    author = "Hagiwara, Kaoru and Okamura, Naotoshi and Senda, Ken-ichi",
    title = "{The earth matter effects in neutrino oscillation experiments from Tokai to Kamioka and Korea}",
    eprint = "1107.5857",
    archivePrefix = "arXiv",
    primaryClass = "hep-ph",
    reportNumber = "KEK-TH1100",
    doi = "10.1007/JHEP09(2011)082",
    journal = "JHEP",
    volume = "09",
    pages = "082",
    year = "2011"
}

@article{Ge:2013zua,
    author = "Ge, Shao-Feng and Hagiwara, Kaoru and Rott, Carsten",
    title = "{A Novel Approach to Study Atmospheric Neutrino Oscillation}",
    eprint = "1309.3176",
    archivePrefix = "arXiv",
    primaryClass = "hep-ph",
    reportNumber = "KEK-TH-1662",
    doi = "10.1007/JHEP06(2014)150",
    journal = "JHEP",
    volume = "06",
    pages = "150",
    year = "2014"
}

@article{Denton:2025oxf,
    author = "Denton, Peter B. and Parke, Stephen J.",
    title = "{Efficient atmospheric, solar, and supernova neutrino propagation through the Earth}",
    eprint = "2511.04735",
    archivePrefix = "arXiv",
    primaryClass = "hep-ph",
    reportNumber = "FERMILAB-PUB-25-0704-T",
    doi = "10.1103/jgw1-lnrl",
    journal = "Phys. Rev. D",
    volume = "113",
    number = "5",
    pages = "053002",
    year = "2026"
}

@article{Krastev:1988yu,
    author = "Krastev, P. I. and Petcov, S. T.",
    title = "{Resonance Amplification and t Violation Effects in Three Neutrino Oscillations in the Earth}",
    doi = "10.1016/0370-2693(88)90404-2",
    journal = "Phys. Lett. B",
    volume = "205",
    pages = "84--92",
    year = "1988"
}

@article{Arafune:1996bt,
    author = "Arafune, Jiro and Sato, Joe",
    title = "{CP and T violation test in neutrino oscillation}",
    eprint = "hep-ph/9607437",
    archivePrefix = "arXiv",
    reportNumber = "ICRR-369-96-20",
    doi = "10.1103/PhysRevD.55.1653",
    journal = "Phys. Rev. D",
    volume = "55",
    pages = "1653--1658",
    year = "1997"
}

@article{Toshev:1991ku,
    author = "Toshev, S.",
    title = "{On T violation in matter neutrino oscillations}",
    doi = "10.1142/S0217732391000464",
    journal = "Mod. Phys. Lett. A",
    volume = "6",
    pages = "455--460",
    year = "1991"
}

@article{Toshev:1989vz,
    author = "Toshev, S.",
    title = "{Maximal $T$ Violation in Matter}",
    reportNumber = "Print-89-0250 (SOFIYA)",
    doi = "10.1016/0370-2693(89)91205-7",
    journal = "Phys. Lett. B",
    volume = "226",
    pages = "335--340",
    year = "1989"
}

@article{Kimura:2007mu,
    author = "Kimura, Keiichi and Takamura, Akira and Yoshikawa, Tadashi",
    title = "{Effect of Leptonic CP Phase in nu(mu) ---{\ensuremath{>}} nu(mu) Oscillations}",
    eprint = "0711.1567",
    archivePrefix = "arXiv",
    primaryClass = "hep-ph",
    doi = "10.1088/1126-6708/2008/03/016",
    journal = "JHEP",
    volume = "03",
    pages = "016",
    year = "2008"
}

@article{Abdullahi:2022fkh,
    author = "Abdullahi, Asli Mohamed and Parke, Stephen J.",
    title = "{Neutrino oscillations in matter using the adjugate of the Hamiltonian}",
    eprint = "2212.12565",
    archivePrefix = "arXiv",
    primaryClass = "hep-ph",
    reportNumber = "FERMILAB-PUB-22-922-T",
    doi = "10.1140/epjc/s10052-024-12885-5",
    journal = "Eur. Phys. J. C",
    volume = "84",
    number = "7",
    pages = "707",
    year = "2024"
}

@article{Cabibbo:1977nk,
    author = "Cabibbo, Nicola",
    title = "{Time Reversal Violation in Neutrino Oscillation}",
    reportNumber = "PAR/LPTHE 77-13",
    doi = "10.1016/0370-2693(78)90132-6",
    journal = "Phys. Lett. B",
    volume = "72",
    pages = "333--335",
    year = "1978"
}

@article{Miura:2001pia,
    author = "Miura, Takahiro and Takasugi, Eiichi and Kuno, Yoshitaka and Yoshimura, Masaki",
    title = "{The Matter effect to T violation at a neutrino factory}",
    eprint = "hep-ph/0102111",
    archivePrefix = "arXiv",
    reportNumber = "OU-HET-376",
    doi = "10.1103/PhysRevD.64.013002",
    journal = "Phys. Rev. D",
    volume = "64",
    pages = "013002",
    year = "2001"
}

@article{Miura:2001pi,
    author = "Miura, Takahiro and Shindou, Tetsuo and Takasugi, Eiichi and Yoshimura, Masaki",
    title = "{The Matter fluctuation effect to T violation at a neutrino factory}",
    eprint = "hep-ph/0106086",
    archivePrefix = "arXiv",
    reportNumber = "OU-HET-389",
    doi = "10.1103/PhysRevD.64.073017",
    journal = "Phys. Rev. D",
    volume = "64",
    pages = "073017",
    year = "2001"
}

@article{Kitano:2024kdv,
    author = "Kitano, Ryuichiro and Sato, Joe and Sugama, Sho",
    title = "{T violation at a future neutrino factory}",
    eprint = "2407.05807",
    archivePrefix = "arXiv",
    primaryClass = "hep-ph",
    reportNumber = "KEK-TH-2638",
    doi = "10.1007/JHEP12(2024)014",
    journal = "JHEP",
    volume = "12",
    pages = "014",
    year = "2024"
}

@article{Nunokawa:2007qh,
    author = "Nunokawa, Hiroshi and Parke, Stephen J. and Valle, Jose W. F.",
    title = "{CP Violation and Neutrino Oscillations}",
    eprint = "0710.0554",
    archivePrefix = "arXiv",
    primaryClass = "hep-ph",
    reportNumber = "FERMILAB-PUB-07-521-T, IFIC-07-58",
    doi = "10.1016/j.ppnp.2007.10.001",
    journal = "Prog. Part. Nucl. Phys.",
    volume = "60",
    pages = "338--402",
    year = "2008"
}

@article{Branco:2011zb,
    author = "Branco, G. C. and Felipe, R. Gonzalez and Joaquim, F. R.",
    title = "{Leptonic CP Violation}",
    eprint = "1111.5332",
    archivePrefix = "arXiv",
    primaryClass = "hep-ph",
    reportNumber = "CERN-PH-TH-2011-289, CFTP-11-020",
    doi = "10.1103/RevModPhys.84.515",
    journal = "Rev. Mod. Phys.",
    volume = "84",
    pages = "515--565",
    year = "2012"
}

@article{Dziewonski:1981xy,
    author = "Dziewonski, A. M. and Anderson, D. L.",
    title = "{Preliminary reference earth model}",
    doi = "10.1016/0031-9201(81)90046-7",
    journal = "Phys. Earth Planet. Interiors",
    volume = "25",
    pages = "297--356",
    year = "1981"
}

@article{ParticleDataGroup:2020ssz,
    author = "Zyla, P. A. and others",
    collaboration = "Particle Data Group",
    title = "{Review of Particle Physics}",
    doi = "10.1093/ptep/ptaa104",
    journal = "PTEP",
    volume = "2020",
    number = "8",
    pages = "083C01",
    year = "2020"
}

@article{DUNE:2020ypp,
    author = "Abi, Babak and others",
    collaboration = "DUNE",
    title = "{Deep Underground Neutrino Experiment (DUNE), Far Detector Technical Design Report, Volume II: DUNE Physics}",
    eprint = "2002.03005",
    archivePrefix = "arXiv",
    primaryClass = "hep-ex",
    reportNumber = "FERMILAB-PUB-20-025-ND, FERMILAB-DESIGN-2020-02",
    doi = "10.2172/1599307",
    month = "2",
    year = "2020"
}

@article{Hyper-Kamiokande:2018ofw,
    author = "Abe, K. and others",
    collaboration = "Hyper-Kamiokande",
    title = "{Hyper-Kamiokande Design Report}",
    eprint = "1805.04163",
    archivePrefix = "arXiv",
    primaryClass = "physics.ins-det",
    month = "5",
    year = "2018"
}

@article{Kimura:2006hy,
    author = "Kimura, Keiichi and Takamura, Akira and Yoshikawa, Tadashi",
    title = "{Measurement of Leptonic CP Phase in nu(mu) ---{\ensuremath{>}} nu(mu) Oscillation}",
    eprint = "hep-ph/0605308",
    archivePrefix = "arXiv",
    month = "5",
    year = "2006"
}

@article{Jarlskog:1985ht,
    author = "Jarlskog, C.",
    title = "{Commutator of the Quark Mass Matrices in the Standard Electroweak Model and a Measure of Maximal CP Nonconservation}",
    reportNumber = "USIP-85-14",
    doi = "10.1103/PhysRevLett.55.1039",
    journal = "Phys. Rev. Lett.",
    volume = "55",
    pages = "1039",
    year = "1985"
}

\end{document}